\documentclass[journal]{IEEEtran}
\usepackage{cite}

\ifCLASSINFOpdf
   \usepackage[pdftex]{graphicx}
   \graphicspath{{./}}
\else
\fi
\usepackage{amsmath}

\usepackage{booktabs}
\usepackage{multirow}

\usepackage{xcolor}
\usepackage[belowskip=-14pt,aboveskip=0pt,font=footnotesize]{caption}
\IEEEaftertitletext{\vspace{-3\baselineskip}}

\begin{document}
\onecolumn
\textcopyright~2019 IEEE.  Personal use of this material is permitted.  Permission from IEEE must be obtained for all other uses, in any current or future media, including reprinting/republishing this material for advertising or promotional purposes, creating new collective works, for resale or redistribution to servers or lists, or reuse of any copyrighted component of this work in other works.

\newpage
\twocolumn

%
\title{A {\color{black} Sub-mm$^3$} Ultrasonic Free-floating Implant for Multi-mote Neural Recording}
%
%
%
\author{Mohammad~Meraj~Ghanbari, David~K.~Piech, Konlin~Shen, Sina~Faraji~Alamouti, Cem~Yalcin, Benjamin~C.~Johnson, Jose~M.~Carmena,  Michel~M.~Maharbiz,~and~Rikky~Muller}

\markboth{\textcopyright~2019 IEEE}
{Ghanbari \MakeLowercase{\textit{et al.}}: A Sub-mm Ultrasonic Free-floating Implant for Multi-mote Neural Recording}
%



\maketitle

\begin{abstract}
A 0.8 mm$^3$ wireless, ultrasonically powered, free-floating neural recording implant is presented. The device is comprised only of a 0.25 mm$^2$ recording IC and a single piezoceramic resonator that is used for both power harvesting and data transmission. Uplink data transmission is performed by {\color{black} analog }amplitude modulation of the ultrasound echo. {\color{black} Using a 1.78 MHz main carrier, \textgreater 35 kbps/mote equivalent uplink data rate is achieved.} A technique to linearize the echo amplitude {\color{black} modulation} is introduced, resulting in \textless 1.2\% static nonlinearity of the received signal over a $\pm$10 mV input range. The IC dissipates 37.7 $\mu$W, while the neural recording front-end consumes 4 $\mu$W and achieves a noise floor of 5.3 $\mu$V$_{rms}$ in a 5 kHz bandwidth. This work improves sub-mm recording mote depth by \textgreater 2.5x, resulting in the highest measured depth/volume ratio by $\sim$3x. Orthogonal subcarrier modulation enables simultaneous operation of multiple implants, using a single-element ultrasound external transducer. Dual-mote simultaneous power up and data transmission is demonstrated at a rate of {\color{black} 7} kS/s at the depth of {\color{black} 50} mm.
\end{abstract}


%
\IEEEpeerreviewmaketitle

\section{Introduction}
%
%
%
%
\IEEEPARstart{U}{ntethered}, wireless neural recording implants are an emerging type of neural interface  \cite{maharbiz2017reliable, seo2016wireless, yeon2018towards, lee2018250mum} that enable improved access to signals valuable for disease diagnosis, closing the loop in stimulation systems, and basic neuroscience research. By their distributed nature, individual wireless devices can precisely target anatomical areas of interest {\color{black} such as} deep brain structures or peripheral nerves. Unlike some wireless devices that sit subcranially on the surface of the brain, wireless devices that target deep structures must strictly minimize size to lessen implantation trauma and long-term tissue scarring \cite{ersen2015chronic} that results in signal quality degradation in chronic neural recording \cite{pothof2017comparison}. Reducing device volume to sub-mm3 scales also enables minimally invasive {\color{black} implantation techniques}, such as {\color{black}catheter-based}, laparoscopic or even injection-based procedures. Designing wireless sub-mm scale implants with centimeters-deep operation {\color{black}ranges presents} power delivery and data transmission {\color{black}challenges}. Furthermore, for concurrent recording from multiple sites, the system should also be able to communicate with a network of such implants.

Multiple designs have been reported recently to address the issues outlined above \cite{seo2016wireless,yeon2018towards,  lee2018250mum}. The smallest free-floating neural recording implant was presented in \cite{lee2018250mum} whose maximum theoretical operation depth does not exceed 6 mm {\color{black} due to high tissue attenuation}, and thus is better suited to epicortical neural recording. Sequential inductive coupling (using an implanted high-Q tertiary coil) was presented in \cite{yeon2018towards} for transcranial power transmission to epicortical free-floating implants at a depth of 20 mm. {\color{black} This technique cannot be extended to deep tissue recording} since the tertiary coil has a large form factor and the implants must be on the same plane (similar to \cite{muller2015minimally, ha2017silicon}). {\color{black} Recently, \cite{lee2019implantable} demonstrated uplink data communication with a network of free-floating implants using a random time-division multiple access (TDMA) protocol and a tertiary coil similar to \cite{yeon2018towards} for power transmission. This implementation is also limited to epicortical recording due to its shared RF link, the limited operation range (1 cm), and uplink data rate (10kbps/device). A frequency division multiple access (FDMA) downlink was proposed in \cite{khalifa2018microbead} to communicate with an ensemble of sub-mm scale neural stimulators. This requires the receiving antenna of each implant to be individually designed and tuned at a unique frequency, complicating the design and cost when scaled to multiple motes.}

{\color{black} }
\begin{figure}[!t]
\centering
\includegraphics[width=1\linewidth]{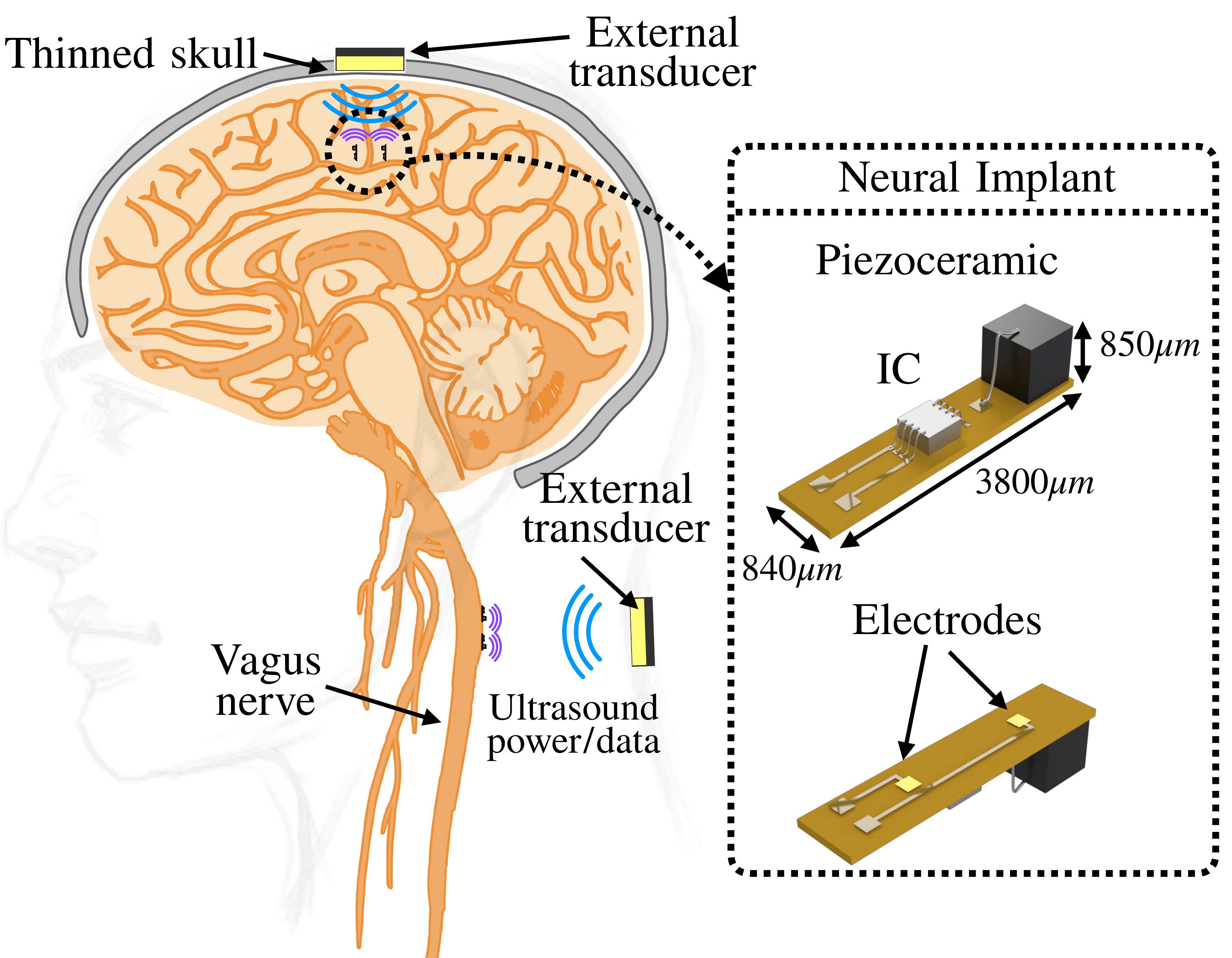}
\caption{Untethered neural recording from deep regions of {\color{black} the peripheral/centra nervous system} using ultrasonically powered neural recording implant.}\label{Fig:motivation}
\end{figure}
\begin{figure*}[!t]
  \includegraphics[width=\textwidth]{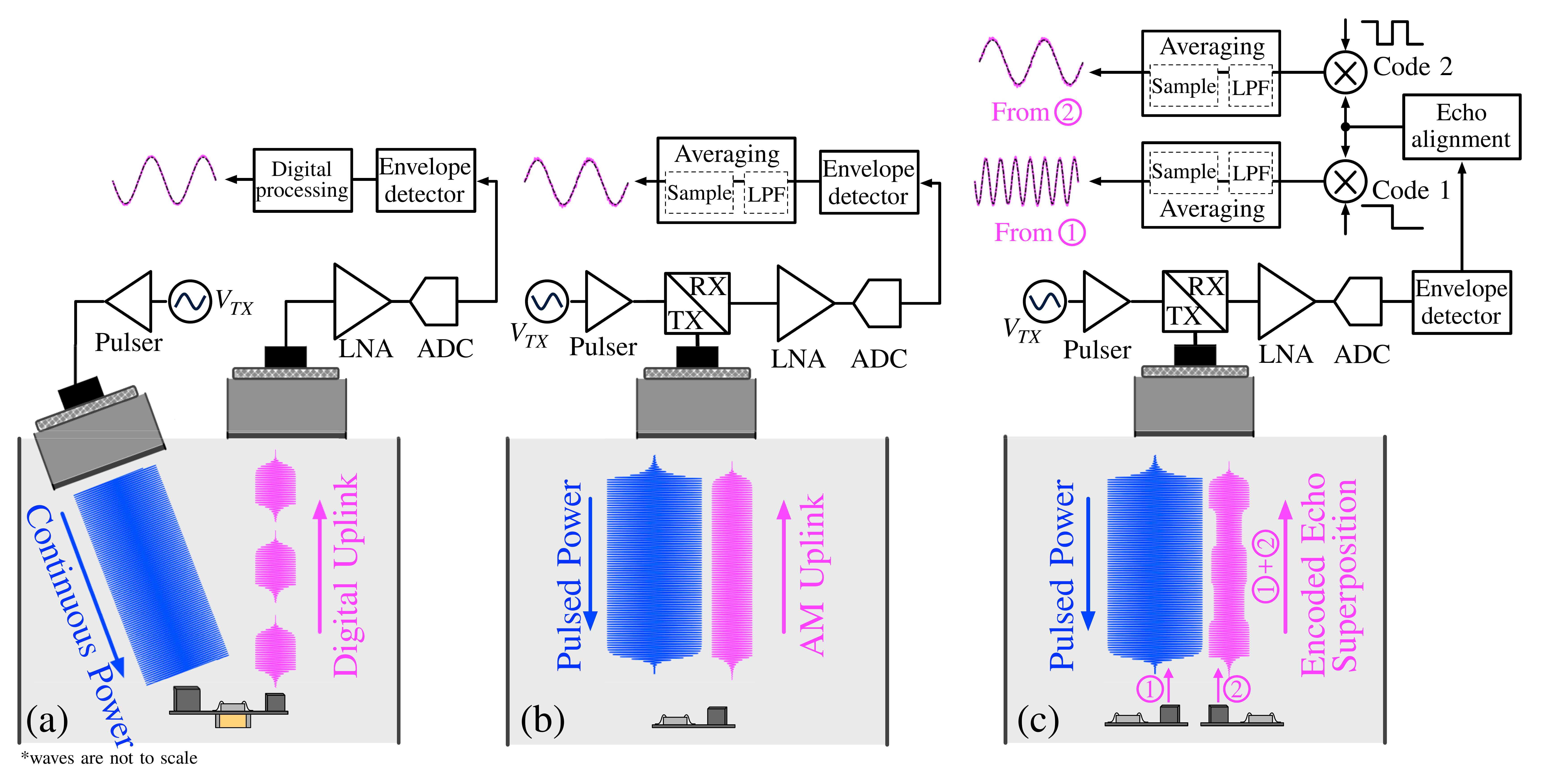}
  \caption{{\color{black}Various ultrasound operating protocols: (a) continuous mode operation where OOK is used for uplink data transmission (b) pulse-echo mode with classic AM uplink data transmission; (c) pulse-echo mode with simultaneous interrogation of two implants using orthogonally encoded AM.}}
  \label{Fig:protocol_comp}
\end{figure*}

The implant presented in \cite{seo2016wireless} takes advantage of low tissue propagation loss ($\sim$0.5 dB/cm/MHz) and low propagation velocity of acoustic waves in realizing a miniaturized, ultrasonically powered implant. However, the absence of a low-noise gain stage limits the SNR of the recorded signal and necessitates the use of a focused ultrasound transducer, restricting tissue operation depth range to 8.8 mm. {\color{black} Electromagnetically powered implants that are \textless 1 mm$^3$ do not meet the depth requirement for recording from the peripheral nervous system (PNS)} in humans targets such as the Vagus nerve, which is located 3 -- 5 cm below the skin surface \cite{hammer2018cervical}. {\color{black} Furthermore, deep-tissue, multi-site neural recording using free-floating implants has not been demonstrated in the aforementioned prior art.} A custom designed beamforming transducer was shown in  \cite{wang2017closed} that can sequentially power up two general-purpose ultrasonically powered implants. However, due to the continuous operating protocol used in  \cite{wang2017closed}, further miniaturization below {\color{black} a} mm$^3$ implant volume is challenging as discussed below.

{\color{black} We present the design, implementation and verification of an ultrasonically powered neural recording implant, shown in Fig. \ref{Fig:motivation}, \cite{ghanbari201917}, that achieves state of the art neural recording performance when compared with other sub-mm$^3$, free floating wireless implants. The implant occupies a volume of only 0.8 mm$^3$, minimizing tissue displacement, scarring, and foreign body response.} The implants have been verified to operate at 50 mm depth {\color{black} in a tissue phantom (with $\sim$ 0.5 dB/cm attenuation at 2 MHz)}, enabling recording of most peripheral nerve targets as well as deep brain targets through thinned skull \cite{fry1978acoustical}. The implants are designed to enable simultaneous power up and parallel data back-telemetry of multiple motes with a low-cost single-element unfocused external transducer. This not only simplifies the design of the external interrogator, but also maximizes operation depth and interrogation frequency (and hence temporal resolution of the acquired signal) in a multi-site recording setup. 

The manuscript is organized as follows. In Section \ref{sec:overview}, design requirements and challenges are outlined. The concept of linear amplitude modulation (AM) of echo for uplink data transmission is introduced in Section \ref{sec:linear_echo} where a theoretical analysis is performed. Section \ref{sec:IC} describes the circuit-level implementation of the integrated circuit (IC) and measurement results are presented in Section \ref{sec:measurement}. Conclusions and comparison to the state of the art are presented in Section \ref{sec:summary}.

\section{System Overview} \label{sec:overview}
To miniaturize wireless implants below millimeter scales, the number and size of off-chip components must be minimized. This includes the elimination of off-chip capacitors and necessitates the realization of wireless power and data communication on a single link. 

Separate power transmission and data communication links have been demonstrated in ultrasonic implants {\color{black}\cite{chang201727, meng2019gastric}}. This mode of operation, shown in Fig. \ref{Fig:protocol_comp}(a), enables continuous data transmission and high data rates, but limits {\color{black}the} miniaturization of {\color{black}the} implant volume since it requires two ultrasound resonators {\color{black}preferably} tuned at distant frequencies to minimize carrier leakage. {\color{black}A similar implant with a single power-data time-multiplexed piezo was presented in \cite{weber2018miniaturized} to reduce the implant volume. However, actively driving the piezo increases the number of off-chip components (storage and matching network capacitors) and ultimately its overall size.} Alternatively, a single ultrasound link can be used for both power and uplink data transmission in a pulse-echo fashion \cite{seo2016wireless}, obviating the need for any secondary resonator or off-chip capacitor. In this scheme, shown in Fig. \ref{Fig:protocol_comp}(b), an ultrasound pulse is first launched towards the implant. After a single time of flight, the implant ultrasound resonator, realized by a bulk piezoceramic (Lead Zirconate Titanate, PZT), starts resonating and harvesting energy. Shortly after that, the integrated circuit (IC) on the implant wakes up and begins recording neural signals. At the same time, the amplitude of the echo (traveling towards the external transducer) is modulated according to the recorded neural signal. The AM-modulated echo is then received and reconstructed through the same external transducer. This pulse-echo interleaved scheme prevents overlapping of the high-voltage signals (up to 30  $\mathrm{V_{peak}}$) driving the external transducer, and the mV-level received echo signals, which would otherwise impose an impractically large dynamic range (e.g. 110 dB and 30 V input range) on the external receiver {\color{black}frontend}.

To minimize the number of off-chip components and the overall implant volume, a pulse-echo interleaved scheme similar to \cite{seo2016wireless} is used in this work, with three key differences. (1) The addition of a low-noise analog front-end in this work reduces the input referred noise by 34x. (2) The introduction of a technique to linearize the reflection coefficient, resulting in linear analog amplitude modulation of the echo and thereby lowering distortion as discussed in section \ref{sec:linear_echo}. (3) Simultaneous multi-implant interrogation is achieved without sacrificing interrogation frequency and with the use of a single-element external transducer. 

{\color{black}Both focal length (Fresnel distance) and focal area of an ultrasound transducer scale with its aperture. For instance, a} low cost commercially available single-element 0.5$''$ diameter unfocused external transducer has a focal {\color{black}length} of 52 mm and focal area of 50 mm$^2$ at 2 MHz in water. Therefore, we propose a network of sub-mm scale implants scattered over this 50 mm$^2$ focal area that simultaneously power up and perform data back telemetry. For uplink data transmission, each implant has a unique orthogonal subcarrier {\color{black}that utilizes} code-division multiplexing (CDM), while modulating the amplitude of its echo. {\color{black}In this prototype, the chip internally generates a CDM code by dividing the clock (extracted from the main carrier) by a ripple counter. In a multi-implant setup CDM codes can be generated in the same fashion \cite{lee2005logic}; a frequency divider is clocked by the extracted global main carrier and followed by an encoder to generate CDM codes. A unique code may be chosen by trimming or hardwiring the device.} The encoded echoes from multiple implants are superimposed in the acoustic medium and received by the external transducer as shown in Fig. \ref{Fig:protocol_comp}(c). {\color{black}The receive chain of the interrogator includes a low noise amplifier and a high-resolution ADC.} Decoding an echo is only a matter of synchronized code multiplication and averaging. {\color{black}Upon echo and CDM code multiplication at the receiver, the signal associated with the CDM code is converted to baseband while those of other channels will remain spread across the spectrum. Averaging concurrently generates a single sample of the selected channel and filters out the non-selected channels. D}ecoding is possible regardless of the length of the encoded echo as long as it contains an instance of a CDM frame. This is crucial because the duration of the time-interleaved echoes is finite (and often short, $\sim$10's of microseconds of microseconds). In addition, orthogonal codes can serve as a subcarrier signal that partially bypasses the low-frequency noise contents of the main carrier \cite{rowe1964amplitude}. 

\begin{figure}[!t]
\centering
\includegraphics[width=1\linewidth]{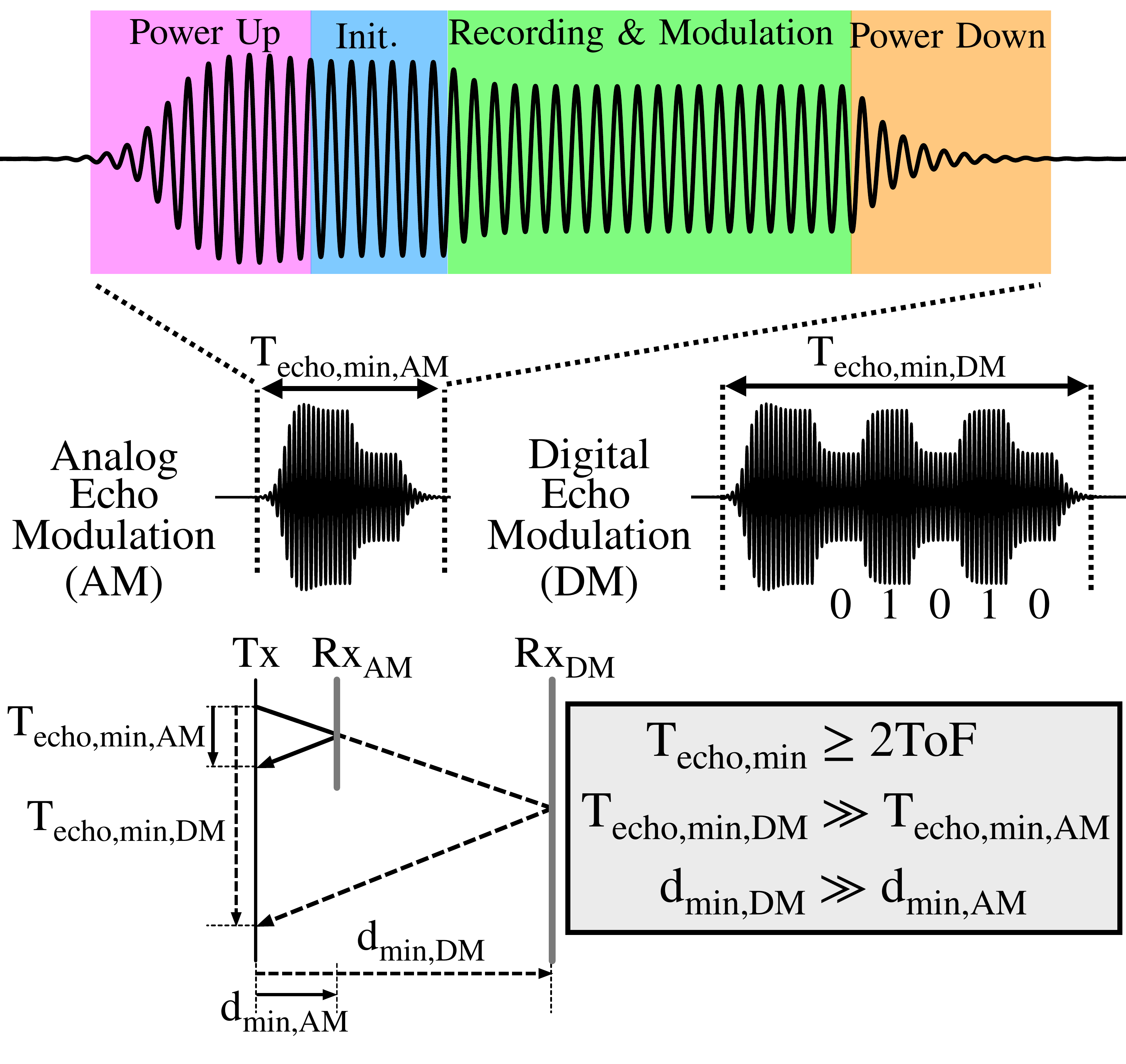}
\caption{{\color{black}(top) A typical ultrasound pulse with analog echo modulation (AM); (bottom) comparison between Analog and Digital echo modulation schemes and their corresponding bounce diagrams}}\label{Fig:analog_vs_digital}
\end{figure}

Fig.  \ref{Fig:analog_vs_digital} shows a typical echo pulse of the implant when {\color{black}analog echo} modulation (AM) is used for data back-telemetry. After implant power up and initialization, the amplitude of the echo is modulated according to the recorded neural signal. A comparison between the required echo period $\mathrm{T_{echo,min}}$ for analog and {\color{black}digital echo} modulation (DM) along with their bounce diagrams are shown in Fig. \ref{Fig:analog_vs_digital}. Assuming the same number of ultrasound cycles ($\mathrm{1/f_{main,carrier}}$) is required for the amplitude to settle or switch between states, the required pulse period for the case of DM is larger than that of AM by roughly the number of digital bits transmitted in each echo. In other words AM carries higher information per cycle than DM. To prevent overlapping the transmitted pulse and the echo, the pulse duration must be smaller than the round-trip time between the external interrogator (Tx) and the implant (Rx), or \textless 2ToF. Thus, the minimum distance between the implant and the external transducer in the case of B-bit digital modulation, $\mathrm{d_{min,DM}}$, is B times larger than that of AM. For a $\mathrm{d_{min,AM}}$ distance shown in Fig. \ref{Fig:analog_vs_digital}, only a single bit of data can be transmitted using DM. The same {\color{black}principle} holds true when subcarrier modulation takes place. {\color{black}The maximum interrogation frequency in a pulse-echo communication channel is given by $\mathrm{f_{sample}=1/2T_{echo,min}}$. In addition to extending the operating range (by allowing shorter distances between Tx and Rx), AM uplink requires shorter $\mathrm{T_{echo,min}}$ and hence can enable higher interrogation frequencies and ultimately uplink data rates.}
\begin{figure}[!t]
\centering
\includegraphics[width=1\linewidth]{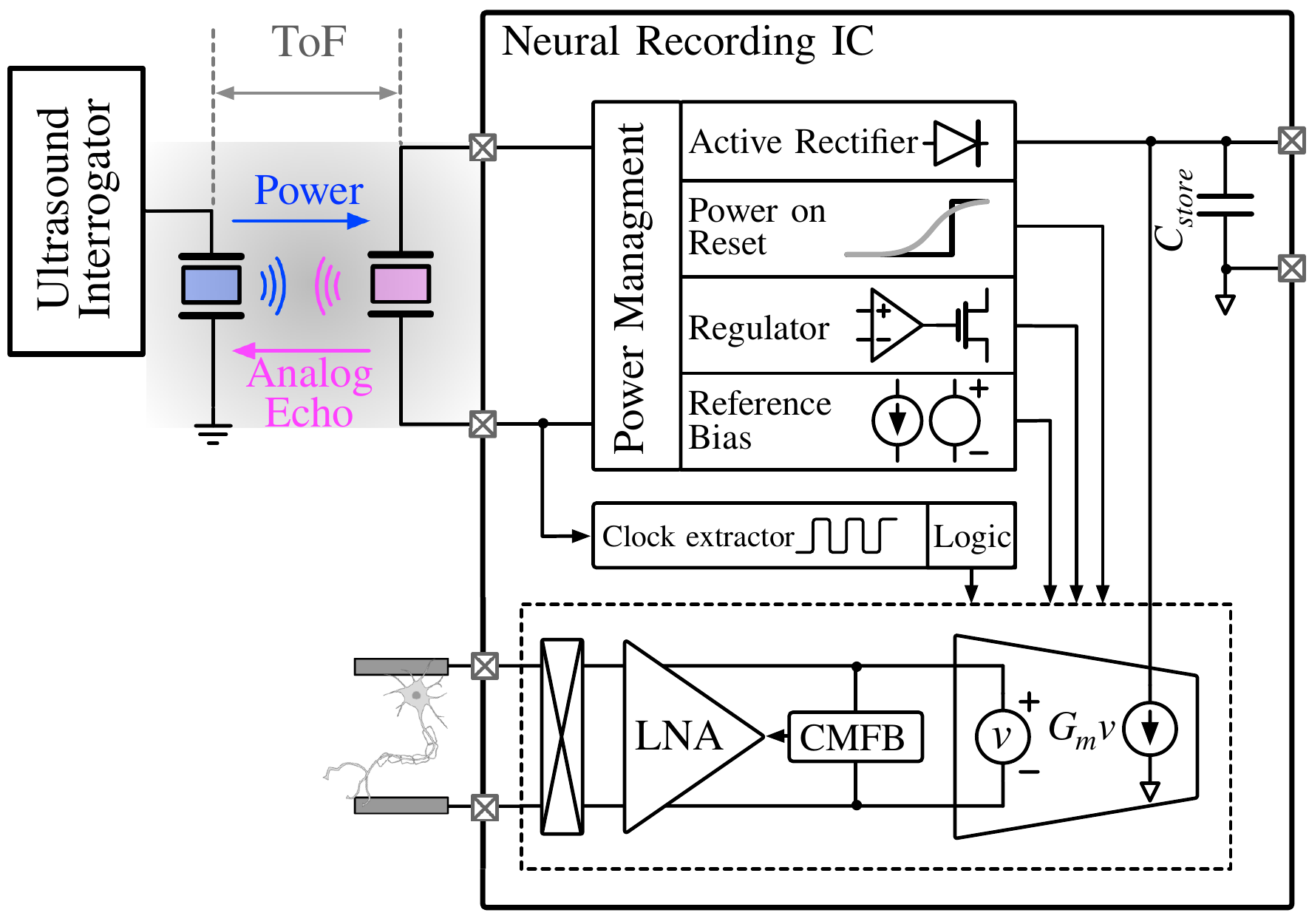}
\caption{Simplified block diagram of the IC.}\label{Fig:diagram}
\end{figure}

A simplified block diagram of the implant is shown in Fig. \ref{Fig:diagram}. The chip contains a power management block that rectifies and regulates the received piezoceramic (piezo) voltage to a 1 V supply. A clock signal is directly extracted from the piezo-harvested voltage {\color{black}(using 2.5 V buffers)}, which is divided to generate the chopper {\color{black}signal}. The analog front-end consists of a fully differential chopper stabilized amplifier followed by a linear gm-cell that linearly modulates the amplitude of the echo. After receiving and conditioning the echo at the external interrogator, {\color{black}digital post-processing}, shown in Fig. \ref{Fig:protocol_comp}(c), is performed to reconstruct the transmitted signal. The implant is powered on as long as an ultrasound pulse is present, meaning that the implant is memory-less. This hinders electrode DC offset cancellation. Therefore, the input-output linear range of the implant should be extended to minimize distortion. The input linear range of the implant is beyond $\pm$10 mV. The output (echo) modulation linearity is achieved by linearly modulating the piezo voltage as discussed in Section \ref{sec:IC}. 

\section{Linear Echo Modulation}  \label{sec:linear_echo}
\begin{figure}[!t]
\centering
\includegraphics[width=1\linewidth]{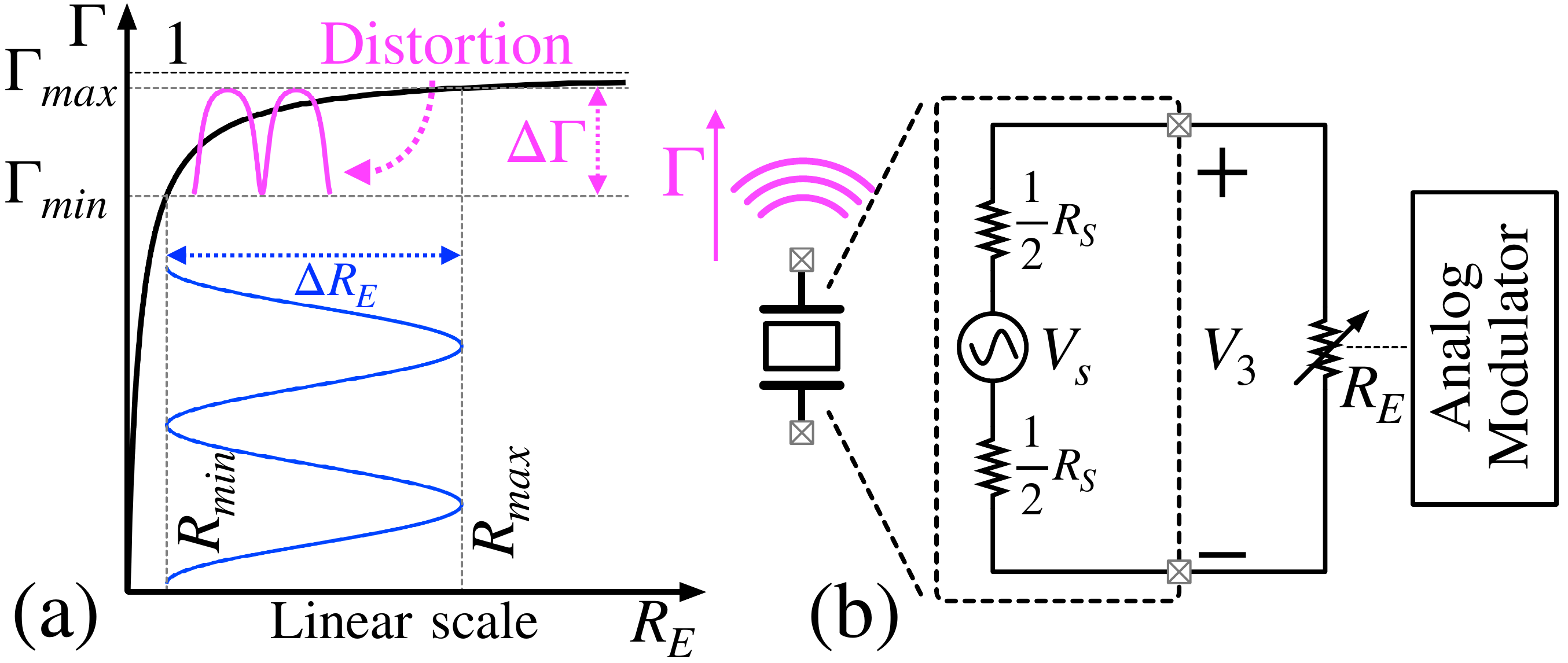}
\caption{(a) Echo amplitude modulation distortion caused by $\Gamma$-$R_E$ nonlinearity (b) piezo equivalent circuit at $f_s$.}\label{Fig:analog_linearity}
\end{figure}
\begin{figure}[!t]
\centering
\includegraphics[width=1\linewidth]{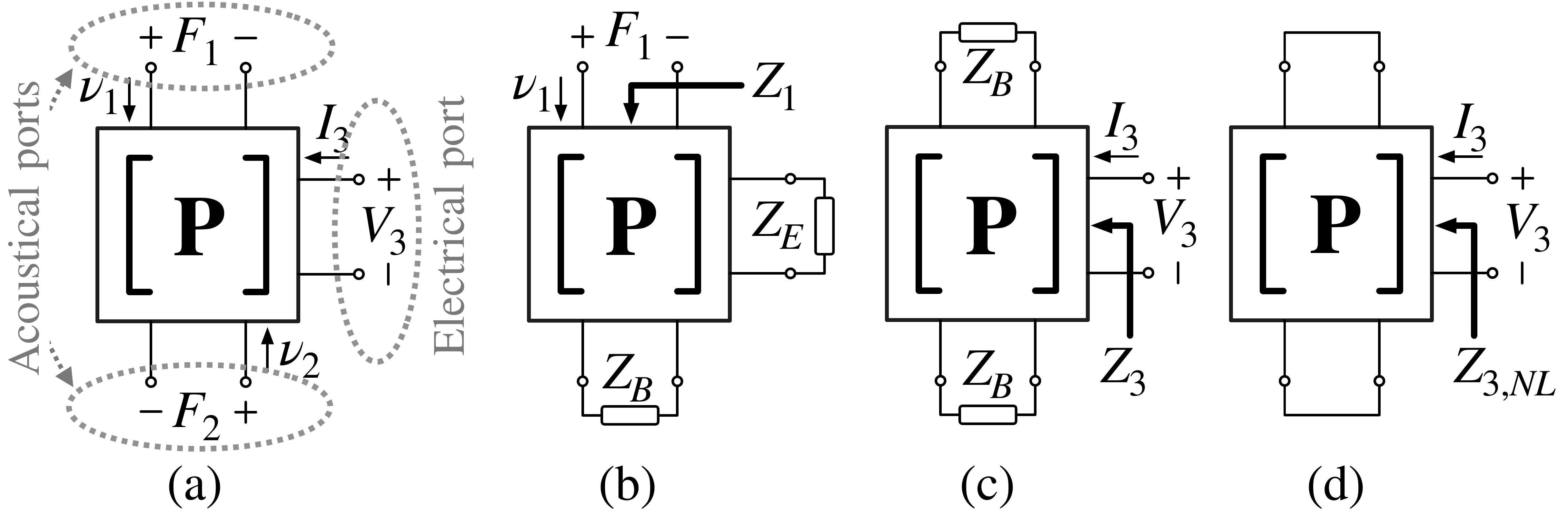}
\caption{(a) Piezo resonator modelled by a three-port network defined by matrix \textbf{P} (b) $Z_1$ is the acoustical impedance seen into port 1 while port 2 is terminated by tissue acoustic impedance $Z_B$ and port 3 is terminated by $Z_E$. (c) $Z_3$ is the electrical impedance seen into port 3 while the acoustical terminals are terminated by $Z_B$ (d) $Z_{3,NL}$ is the unloaded electrical impedance of the piezo.}\label{Fig:matrix}
\end{figure}

The profile of the acoustic reflection coefficient $\Gamma$ of a piezo as a function of its termination resistance $R_E$ is shown in Fig. \ref{Fig:analog_linearity}(a). It can be observed that the modulated echo is significantly distorted due to  {\color{black}the} nonlinearity of $\Gamma$, especially when concurrent energy harvesting with echo modulation imposes a minimum value of termination resistance  $R_E = R_{min}$. That is, $R_{min}$ should be large enough to allow energy harvesting. This is not a  {\color{black}typical} problem for digital modulation when transmission of only two states is  {\color{black}needed}. For AM modulation, however, the source of this nonlinearity should be understood and if possible, linearized through co-design of the piezo and the modulating IC. However, the governing equations of bulk piezos as well as their equivalent circuit models (KLM \cite{krimholtz1970new} and Redwood \cite{redwood1961transient}) are complex  {\color{black}and provide little} insight into the source of this nonlinearity. Instead, analytical derivation and experimental verification of a simple expression can guide the modulator design and lead to linear echo modulation received at the external interrogator.  {\color{black}Such an expression is introduced in this section and} used in Section \ref{sec:IC} to implement a linear echo amplitude modulator.
\begin{figure}[!t]
\centering
\includegraphics[width=1\linewidth]{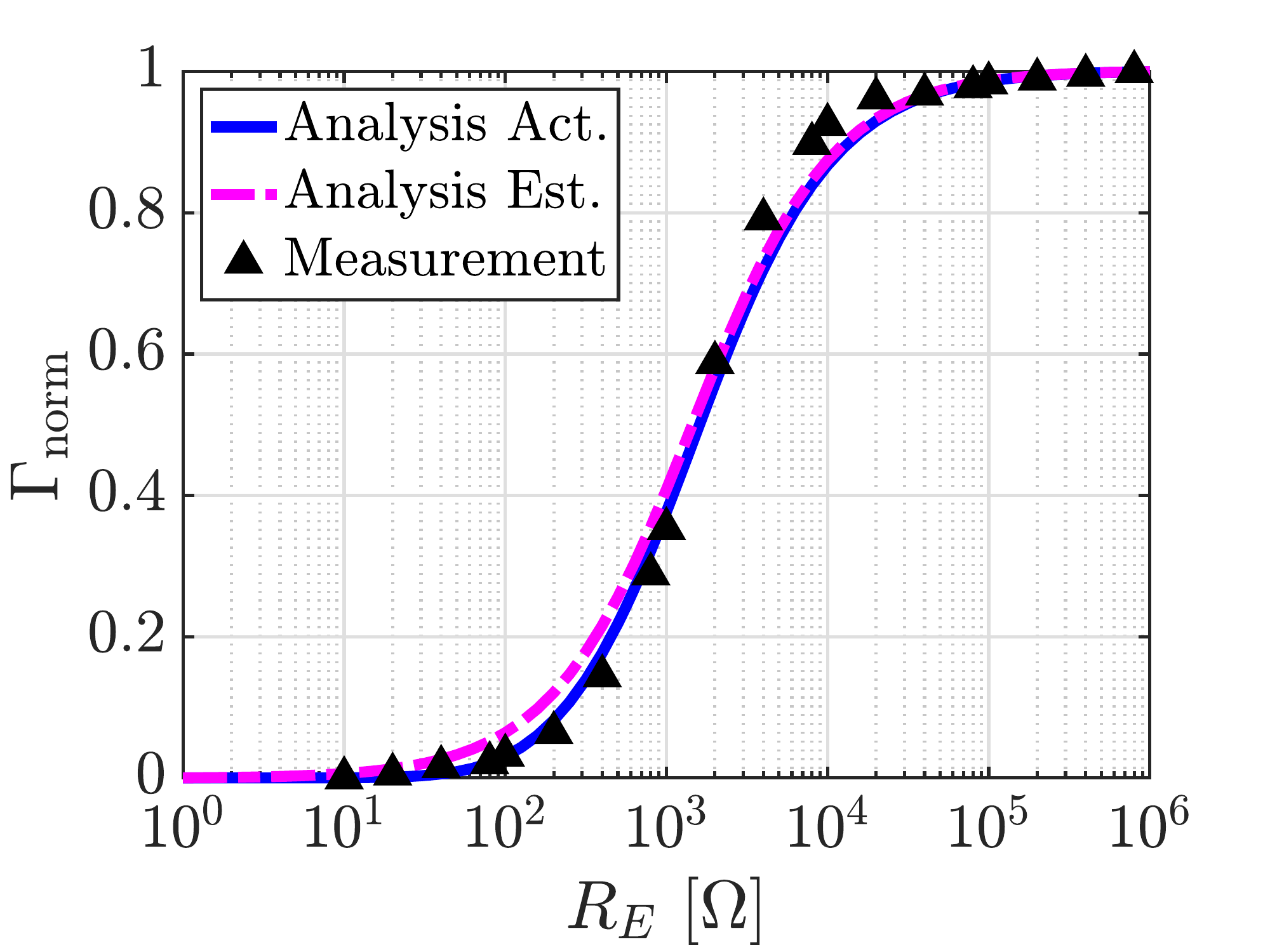}
\caption{Normalized $\Gamma$-$R_E$ curve obtained by (\ref{eq:gamma_def3}), (\ref{eq:gamma_approx_33}) and measurement.}\label{Fig:gamma_vs_RE}
\end{figure}
The piezo is modeled as a thickness-mode resonating 3-port network, shown in Fig. \ref{Fig:matrix}(a), whose input-output port relationships are well described by \cite{kino1987acoustic},
\begin{IEEEeqnarray}{rclll}
	\begin{bmatrix}
    F_1 \\
    F_2 \\
    V_3 \\
   \end{bmatrix} &=& \mathbf{P}\begin{bmatrix}
    \nu_1 \\
    \nu_2 \\
    I_3 \\
   \end{bmatrix}&=& 
	   	\begin{bmatrix}
   m & n & p \\
   n & m & p \\
   p & p & r \\
   \end{bmatrix}
   	\begin{bmatrix}
    \nu_1 \\
    \nu_2 \\
    I_3 \\
   \end{bmatrix}
\end{IEEEeqnarray}
\begin{IEEEeqnarray}{rcl}
\begin{bmatrix}
    F_1 \\
    F_2 \\
    V_3 \\
   \end{bmatrix} &=&
   \begin{bmatrix}
    \frac{Z_0A}{j\mathrm{tan}(\beta l)} & \frac{Z_0A}{j\mathrm{sin}(\beta l)} & \frac{h_{33}}{j\omega} \\
    \frac{Z_0A}{j\mathrm{tan}(\beta l)} & \frac{Z_0A}{j\mathrm{sin}(\beta l)} & \frac{h_{33}}{j\omega} \\
    \frac{h_{33}}{j\omega} & \frac{h_{33}}{j\omega}& \frac{1}{j\omega C_0} \\
   \end{bmatrix}
   \begin{bmatrix}
    \nu_1 \\
    \nu_2 \\
    I_3 \\
   \end{bmatrix}
   \label{eq:first3}
\end{IEEEeqnarray}
Ports 1 and 2 are acoustical, and port 3 is the electrical port of the piezo. Table \ref{Tab:param} describes the parameters used in (\ref{eq:first3}). The acoustic impedance seen into port 1, while port 2 and 3 are respectively terminated by $Z_B$ and $Z_E$ (Fig. \ref{Fig:matrix}(b)) is given by
\begin{IEEEeqnarray}{rcl}
Z_1 &=& \frac{p^2(2n-2m-Z_B)+(Z_E+r)(m^2-n^2+mZ_B)}{(Z_E+r)(m+Z_B)-p^2}.
\end{IEEEeqnarray}
Therefore, $\Gamma$ is given by
\begin{IEEEeqnarray}{rcl}
\Gamma & = & \frac{Z_1-Z_B}{Z_1+Z_B}.
\label{eq:gamma_def3}
\end{IEEEeqnarray}
{\color{black}It is shown in the Appendix that at the series resonance frequency of the piezo, (\ref{eq:gamma_def3}) further simplifies to}
 \begin{IEEEeqnarray}{rclll}
\Gamma &\approx & \frac{Z_E}{Z_E+R_{S}}&\propto &V_3,
\label{eq:gamma_approx_33}
\end{IEEEeqnarray}
 \begin{table}[t]
\renewcommand{\arraystretch}{0.9}
\captionsetup{}
\caption{List of piezoelectric typical parameters}
\label{Tab:param}
\label{table:SC_SAR}
\centering

\begin{tabular}{@{}llll@{}}
\toprule
Parameter  & Unit & Value & Description\\
\midrule
$\rho$ & $\mathrm{kg/m^{3}}$ & $7600$  & Piezo Density \\
$\epsilon_{33}$ & $\mathrm{nF/m}$ & $16.8$ & Piezo Dielectric constant \\
$c_{33}^E$ & $\mathrm{GPa}$ & $50$ & Piezo Elastic constant \\
$c_{33}^D = c_{33}^E(1+k^2)$ & $\mathrm{GPa}$ & $73$ & Stiffened $c_{33}$ \\
$e_{33}$ & $\mathrm{C/m^{2}}$ & $20$ & Piezo Stress constant \\
$h_{33} = e_{33}/\epsilon_{33}$ & $\mathrm{GC/m.F}$ & $1.18$ & Piezo constant \\
$k = e_{33}/\sqrt{\epsilon_{33}.c_{33}^E}$ & -- & $0.69$ & Electromechanical coupling \\
$k_t =  \sqrt{k^2/(1+k^2)}$ & -- & $0.56$ & -- \\
$v_a = \sqrt{c_{33}^E/\rho}$ & $\mathrm{m/s}$ & $2564$ & Piezo wave velocity \\
$\overline{v_a} = v_a\sqrt{1+k^2}$ & $\mathrm{m/s}$ & $3115$ & Stiffened $v_a$ \\
$A$ & $\mathrm{mm^{2}}$ & $0.56$ &  Piezo cross-section area \\
$C_o =  A\epsilon_{33}/t$ & $\mathrm{pF}$ & $12.6$ & Piezo capacitance \\
$Z_o =  \rho \overline{v_a}$ & M$\mathrm{Pa.s/m}$ & $23$ & Piezo acoustic impedance \\
$f_p =  \frac{1}{2}\overline{v_a}/t$ & $\mathrm{MHz}$ & $2.07$ &  Parallel resonance frequency \\
$f_s =  fp/\sqrt{1+8(k/\pi)^2}$ & $\mathrm{MHz}$ & $1.76$ &  Series resonance frequency \\
$\beta =  2\pi f/\overline{v_a}$ & $\mathrm{m^{-1}}$ & -- & wave propagation constant \\

\bottomrule
\end{tabular}
\label{tab:final_perf}

\end{table}
where
\begin{IEEEeqnarray}{rclll}
R_S &=&\frac{-2Z_Bp^2}{(n+m)^2} = 2Z_B(\frac{mr-p^2}{m^2-n^2}),
\end{IEEEeqnarray}
 is the internal series resistance of the piezo. At $f_s$, $\Gamma$ is approximately linearly proportional to the voltage across port 3 (coupling the simplified circuit model of Fig. \ref{Fig:analog_linearity}(b) to the acoustical port of the piezo). Therefore, to linearly modulate $\Gamma$, the voltage across the piezo should be linearly modulated. Fig. \ref{Fig:gamma_vs_RE} compares analytical expression (\ref{eq:gamma_def3}) and its approximation (\ref{eq:gamma_approx_33}) for the parameter values listed in Table I, showing excellent matching between the expressions and the measured values. In contrast to (\ref{eq:gamma_def3}), (\ref{eq:gamma_approx_33}) has a single parameter, $R_S$, which can be obtained empirically or by finite element model (FEM) simulation. To verify the model, measurements were made on a 0.75x0.75x0.75 mm$^3$ piezo (854, APC International), whose $R_S = 1.5 \mathrm{k\Omega}$.

\section{Integrated Circuit Implementation} \label{sec:IC}
\begin{figure}[!t]
\centering
\includegraphics[width=1\linewidth]{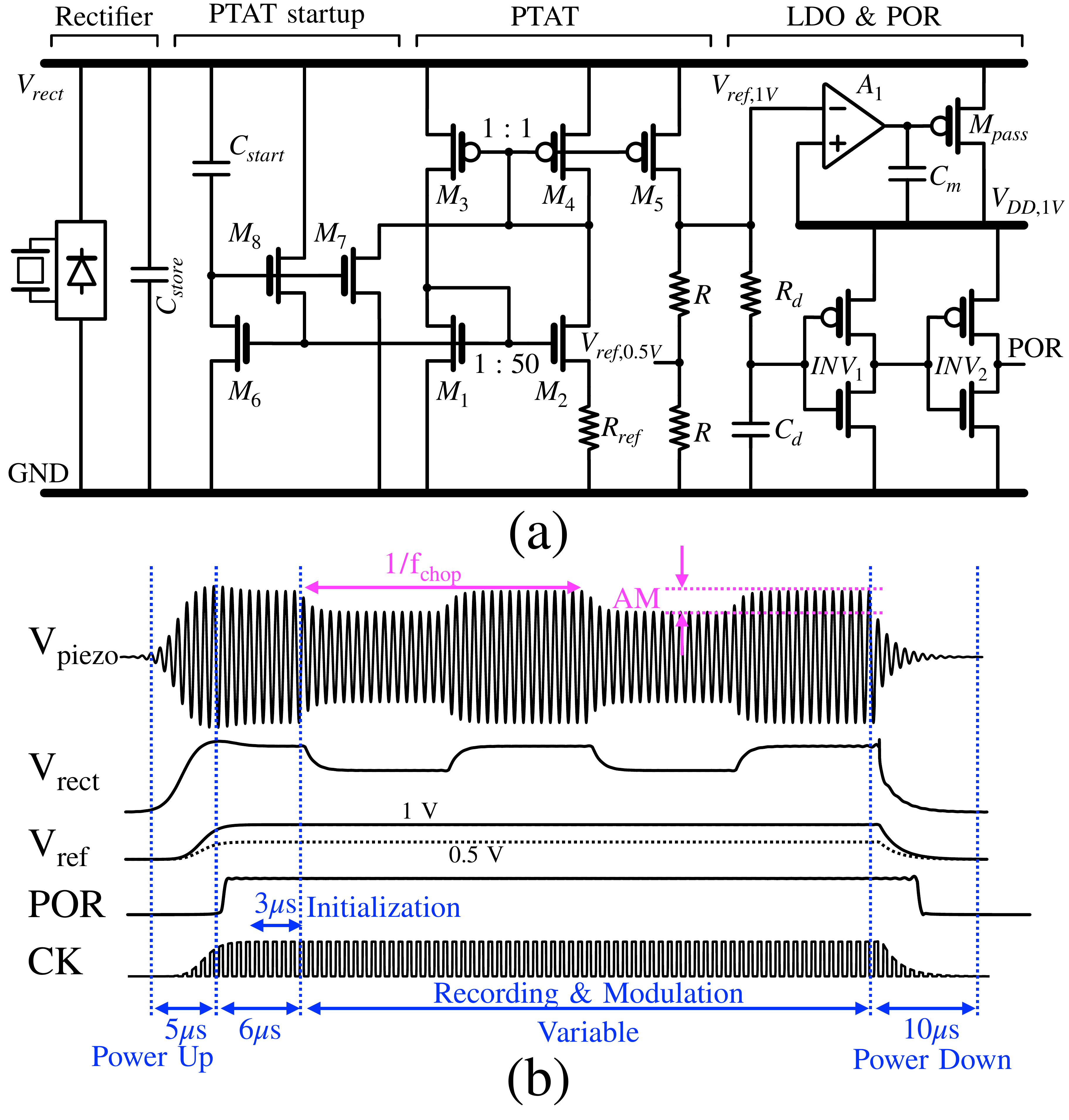}
\caption{ {\color{black}(a) Schematic diagram of power management blocks (b) chip-level timing diagram shown with 2 subcarrier cycles.}}\label{Fig:chip_timing}
\end{figure}

The implemented mote utilizes a 0.75x0.75x0.75 mm$^3$ piezo (840, APC International), whose $R_S = 4 \mathrm{~k\Omega}$. When an ultrasound pulse is received at the piezo, the IC has a finite amount of time (\textless 2ToF, e.g. 66 $\mu$s @ 50 mm depth) to power up, generate a stable supply, record neural signals and perform uplink data transmission. Therefore, rapid power-up and precise timing management of the sub-blocks are critical. The {\color{black}interconnection of power management blocks and the} top-level timing diagram of the chip are shown in Fig. \ref{Fig:chip_timing}. Due to the time constant associated with the on-chip storage capacitor (130 pF), the internal series resistance of the piezo{\color{black}, and the} \textless 2$\pi$ conduction angle of the rectifier, it takes $\sim$5 $\mu$s ($\sim$10 ultrasound cycles) to fully charge up {\color{black}the storage capacitor and for the supply-independent proportional-to-absolute-temperature (PTAT) source to generate two reference voltages ($V_{ref,1V}$ and $V_{ref,0.5V}$) on-chip. The PTAT core transistors M$_{1-4}$ are designed to operate in the subthreshold region such that $I_{M5}=\eta V_T \mathrm{Ln}(W_2/W_1)/R_{ref}$, where $\eta$ and $V_T$ are the subthreshold factor and thermal voltage, respectively. Initially, M$_{1-8}$ are off and the gates of M$_{7-8}$ track $V_{rect}$. Upon harvesting voltage from the piezo, $V_{rect}$ rises, and M$_{7-8}$ turn on pulling up/down the gates of M$_{1-2}$/M$_{3-4}$ to speed up their transition from zero current to the desired current (500 nA) stable bias point. M$_6$ is designed to be strong enough to pull down the gates of M$_{7-8}$, charging $C_{start}$ and disengaging the PTAT startup circuitry once M$_{1-2}$ turn on. $V_{ref,1V}$ serves as the low dropout regulator (LDO) reference voltage, and a delayed version of $V_{ref,1V}$ triggers the} power on reset (POR) to initialize the logic states. The amplifier initialization takes 3 $\mu$s, which is followed by signal acquisition and uplink data transmission. In the absence of an ultrasound pulse, the on-chip storage capacitor discharges in $\sim$10 $\mu$s, meaning that inter-pulse duration should be greater than 10 $\mu$s for proper re-initialization (POR triggering) of the IC. This translates to a minimum operation depth of 14 mm without requiring any acoustic spacer. 
\subsection{Low-noise Amplifier}
\begin{figure}[!t]
\centering
\includegraphics[width=1\linewidth]{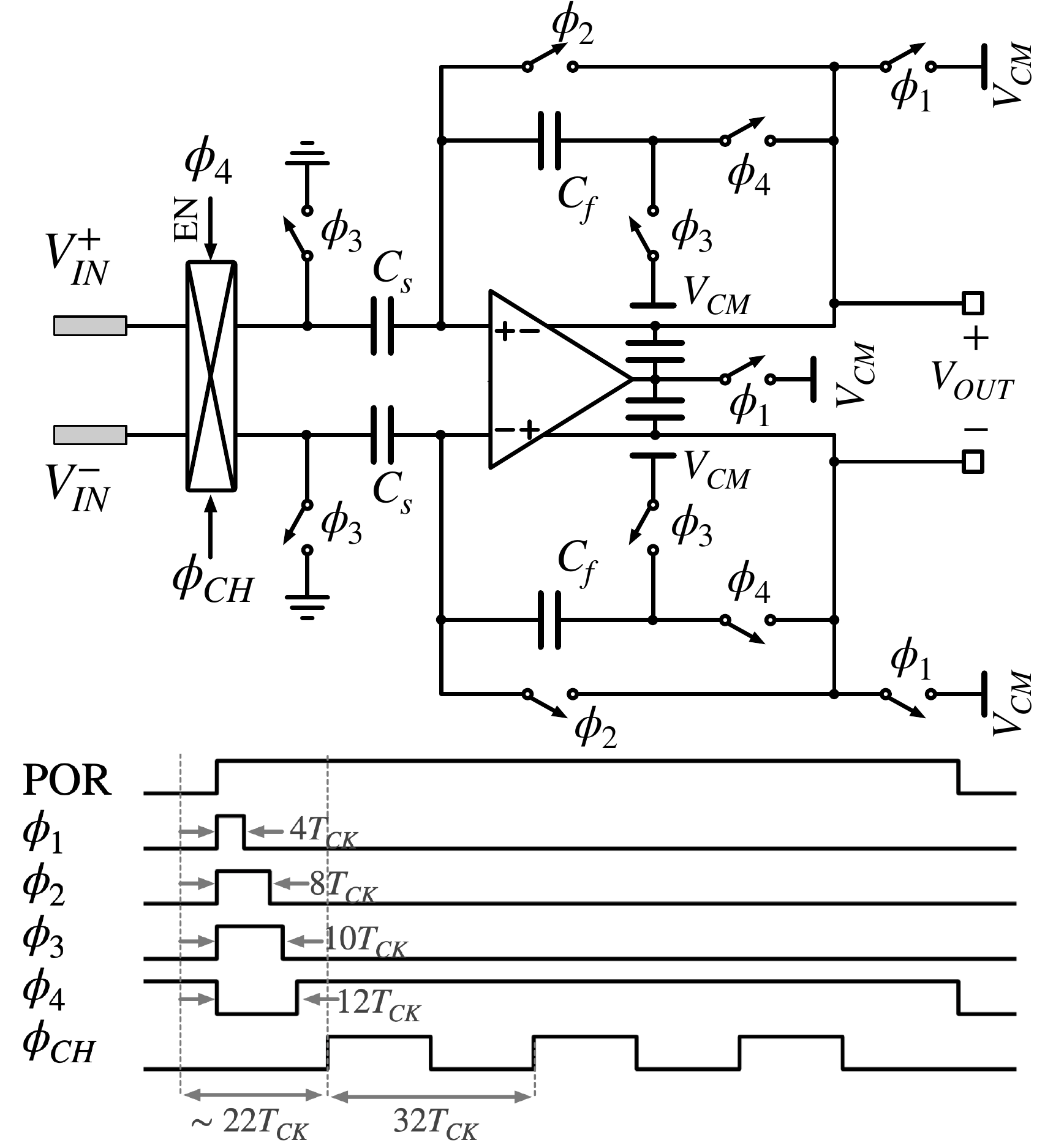}
\caption{LNA topology and timing diagram.}\label{Fig:LNA}
\end{figure}
\begin{figure}[!t]
\centering
\includegraphics[width=1\linewidth]{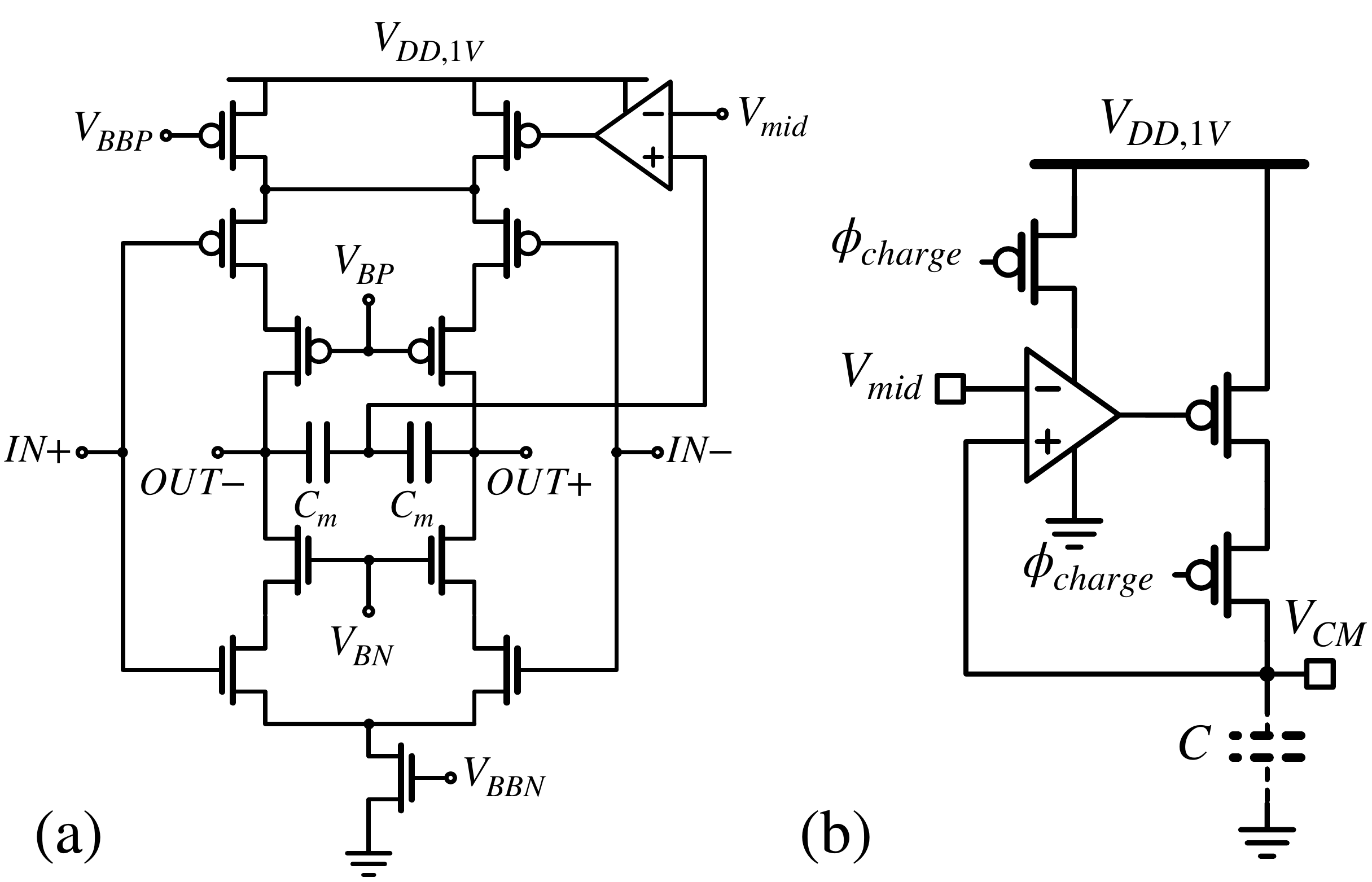}
\caption{(a) Fully complimentary OTA topology used in the LNA (b) power-gated charger of the input terminals of the OTA.}\label{Fig:OTA}
\end{figure}
The analog front-end (AFE) of the chip consists of a fully differential capacitive-feedback low noise amplifier (LNA) whose circuit diagram is shown in Fig. \ref{Fig:LNA}. The values of the feedback and load capacitors (0.44 pF and 4.7 pF respectively), and the feedback factor ($\beta\sim$1/16) set the noise (\textless 10 $\mu\mathrm{V_{rms}}$ in 180 kHz bandwidth) of the circuit, while the effective transconductance of the LNA sets the bandwidth. The amplifier is initialized by a set of switches controlled by $ϕ\phi_{1-4}$, which bias the LNA as quickly as possible to maximize the signal acquisition period. $V_{CM}$ is set to the 0.5 V mid-rail voltage using a power-gated linear regulator shown in Fig. \ref{Fig:OTA}. The low impedance source charges 25 pF of capacitance in \textless 1 $\mu$s. 

Since the LNA runs on a 1 V supply and has a gain of 16, auto-zeroing is implemented to cancel amplifier offset and improve the linear output voltage swing. Auto-zeroing is implemented by sampling the offset of the amplifier on the feedback capacitors ($C_f$). The $\phi_1$ switches are disabled first, placing the amplifier in the unity gain feedback. The sampled offset is then subtracted from the signal after sequentially opening $\phi_2$ and $\phi_3$ and establishing the signal path through $\phi_4$. The sequential switching results in kT/C noise that is added to the sample after the initialization period of each interrogation event. To mitigate kT/C noise, the input signal is chopper stabilized and upmodulated to a frequency $\mathrm{f_{chop}}$ while the kT/C noise remains at baseband. When the signal is downmodulated, the kT/C noise is converted into out-of-band chopper ripple \cite{johnson2017implantable}. Since downmodulation occurs on the interrogator side, chopper stabilization is used to simultaneously bypass the 1/f and kT/C noises of the amplifying circuits as well as the low-frequency noise contents of the main carrier. In presence of chopping switches, the input impedance of the amplifier is given by $Z_{in}=1/(2f_{chop} C_s )\sim 1.3 ~\mathrm{M\Omega}$ at the highest chopping frequency of 55 kHz. {\color{black}The impedance of the electrodes submerged in saline was measured to be an order of magnitude smaller than the input impedance of the AFE (\textless 100 k$\Omega$) for frequencies greater than 100 Hz (Fig. \ref{Fig:meas_electrode}).}

The total output referred noise power of the amplifier is given by $\overline{v_n^2}=kT\alpha\gamma/(\beta C_T )$ \cite{murmann2012thermal}, where $k$ is the Boltzmann constant, $T$ is the absolute temperature, $\alpha$ is the excess noise factor of the OTA, $\gamma$ is the MOS channel noise coefficient, and $C_T$ is the total capacitance seen at the output node during amplification. The bandwidth of the LNA is given by $\omega_{3dB}=\beta G_m/C_T$, where $G_m$ is the effective transconductance of the OTA. A fully complementary differential amplifier topology \cite{bazes1991two} shown in Fig. \ref{Fig:OTA}(a) is chosen since its $\alpha$ is close to 1, and it has a high $G_m/I_D$, since $G_m=g_{m,N}+g_{m,P}$. The output range of the LNA is $\pm$160 mV, with a bandwidth of 180 kHz, high enough to pass the third harmonic of the highest subcarrier frequency ({\color{black} 55 kHz). Although the amplifier has a broadband forward path of 180 kHz, the bandwidth of the post-processed signal, and therefore the effective noise bandwidth of the amplifier are reduced to 5 kHz at an interrogation frequency of 10 kHz. Along the signal chain, as shown in Fig. \ref{Fig:protocol_comp}(c), decoding and demodulation of the echo involves averaging the echo for the duration of the pulse. This averaging concurrently applies a sinc low pass filter with a 3dB bandwidth of $1/2T_{integration}$ to the received signal and translates every received echo to a single sample. Therefore, at a 10 kHz interrogation frequency (sampling frequency), the signal and noise bandwidth is reduced to 5 kHz.}


\subsection{Linear $G_m$-cell design}
\begin{figure}[!t]
\centering
\includegraphics[width=1\linewidth]{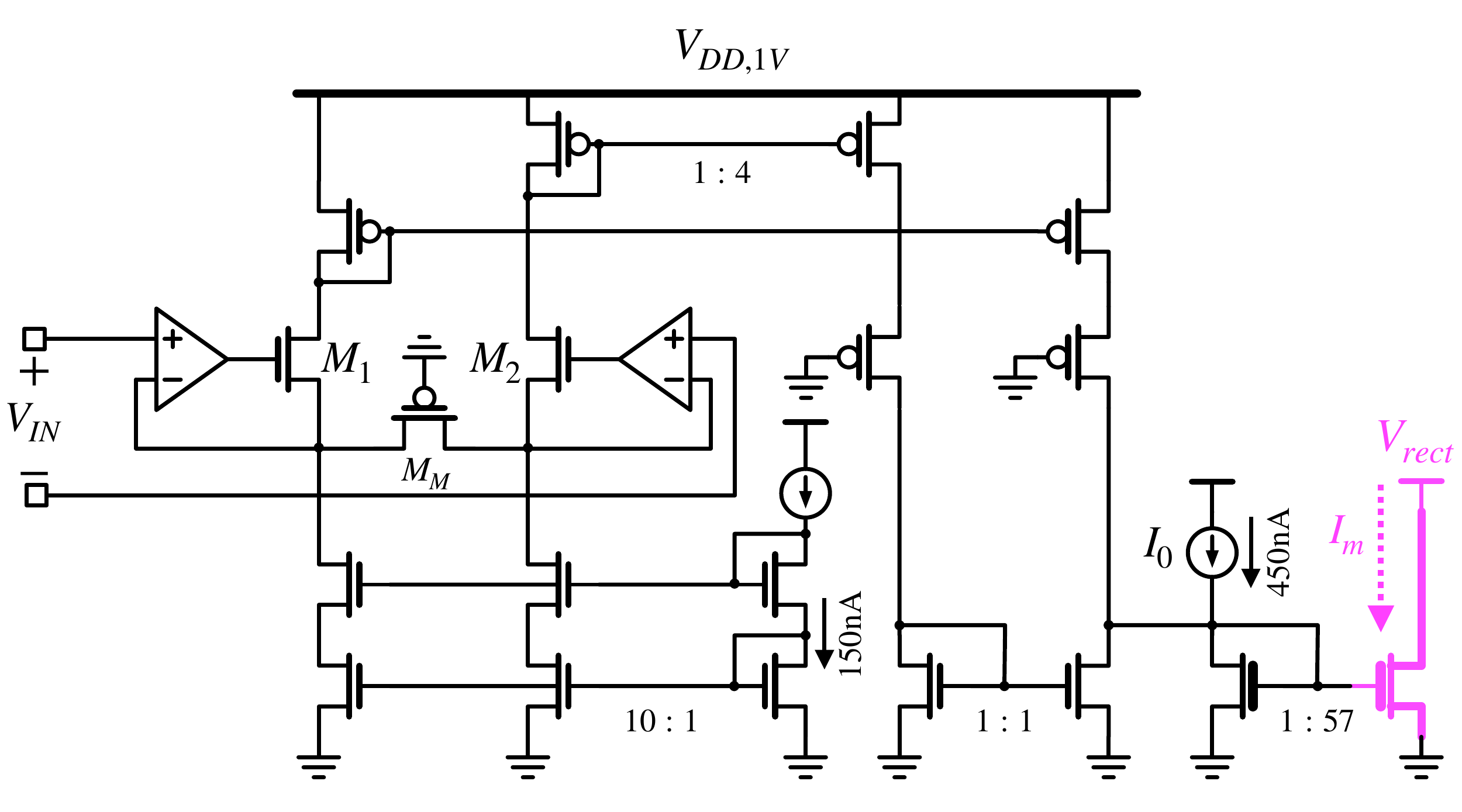}
\caption{Linear gm-cell architecture.}\label{Fig:linera_gm}
\end{figure}

\begin{figure}[!t]
\centering
\includegraphics[width=1\linewidth]{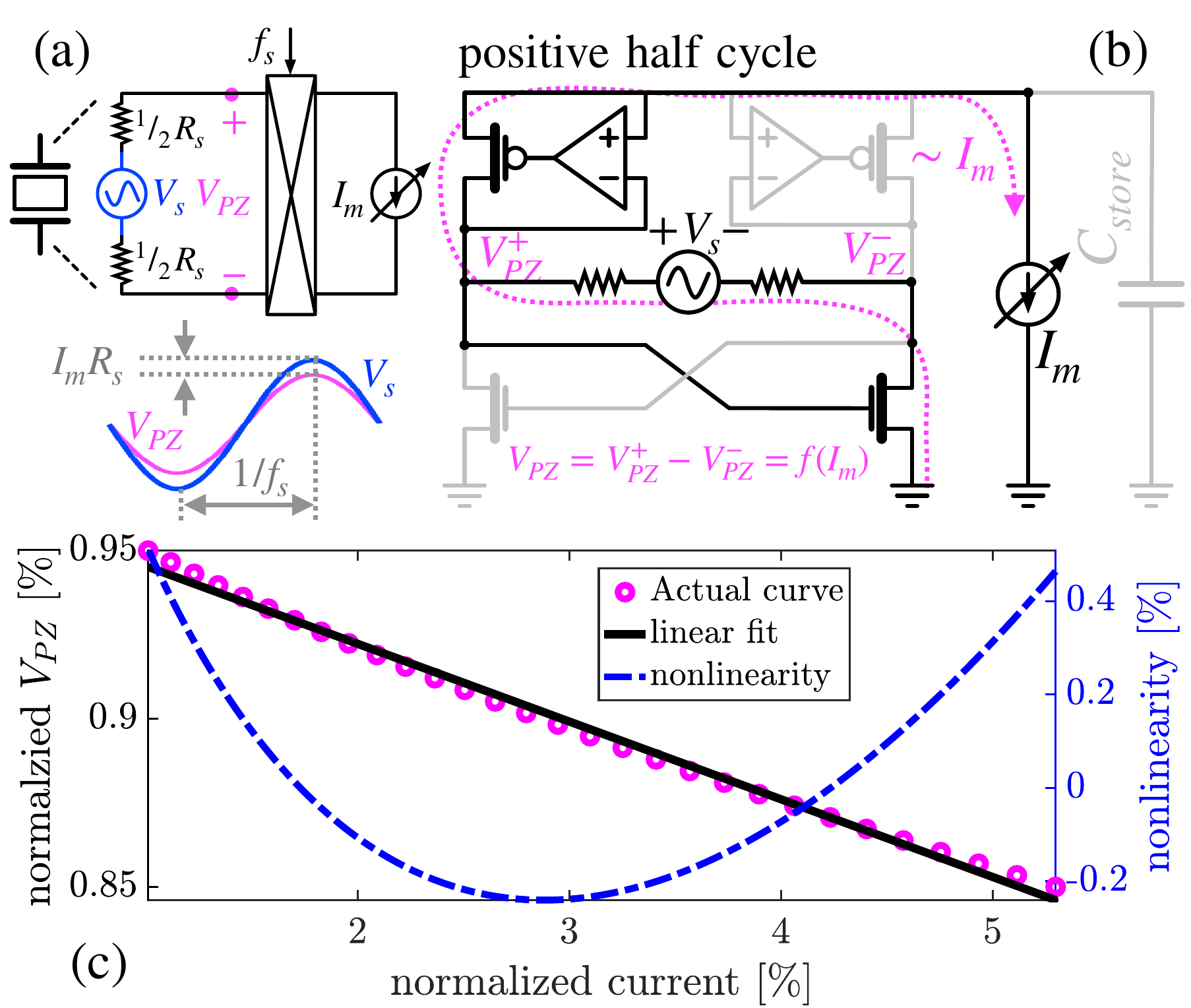}
\caption{(a) Conceptual echo amplitude modulation using synchronous up-conversion current mixer (b) reusing active rectifier as synchronous mixer (c) nonlinearity induced by the rectifying modulator.}\label{Fig:rectifier_nonlinearity}
\end{figure}
\begin{figure}[tb]
\centering
\includegraphics[width=1\linewidth]{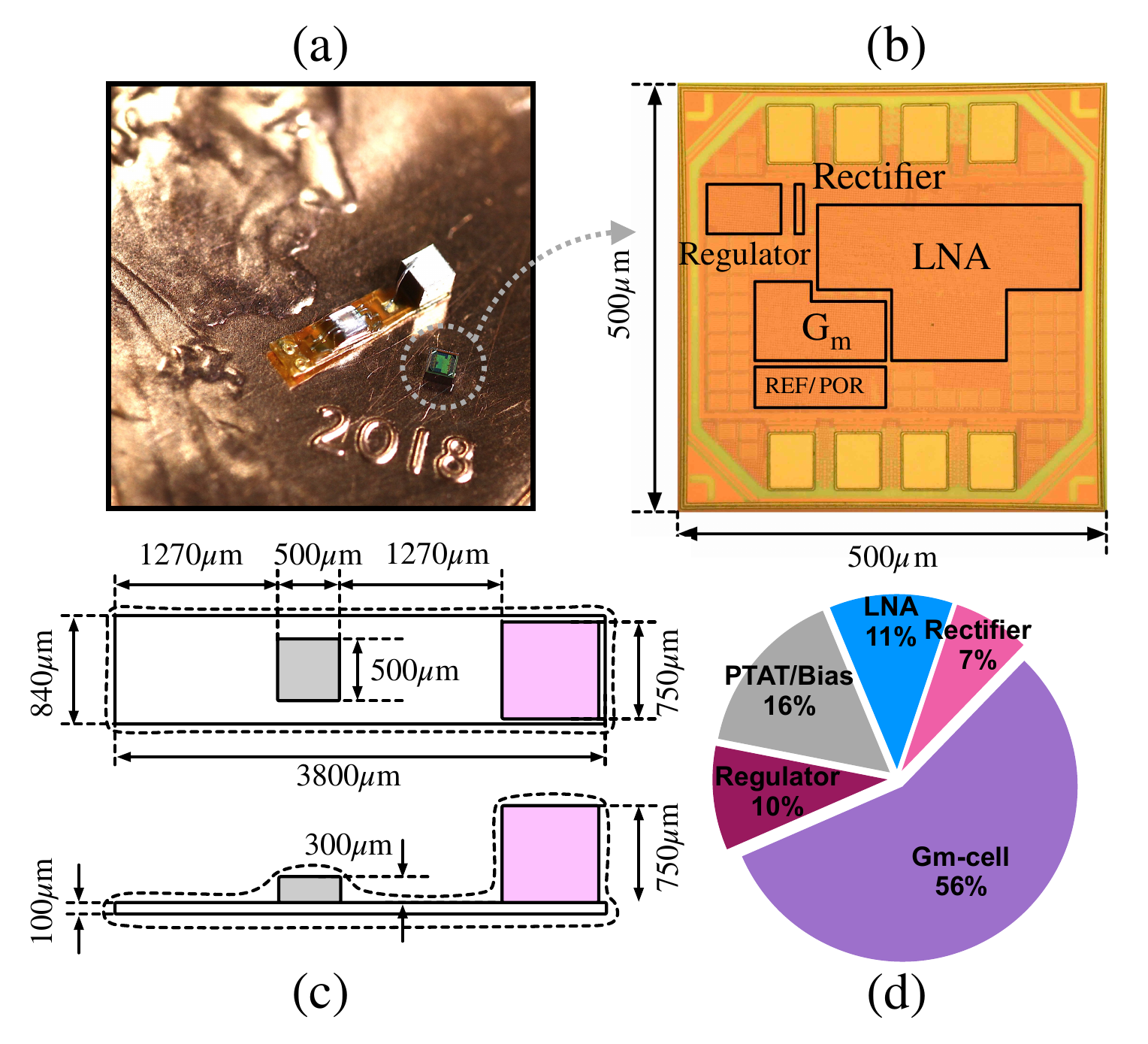}
\caption{(a) Fully packaged implant micrograph (b) IC micrograph (c) dimensions of the fully packaged implant (d) power consumption breakdown.}\label{Fig:chip}
\end{figure}
\begin{figure}[h]
\centering
\includegraphics[width=1\linewidth]{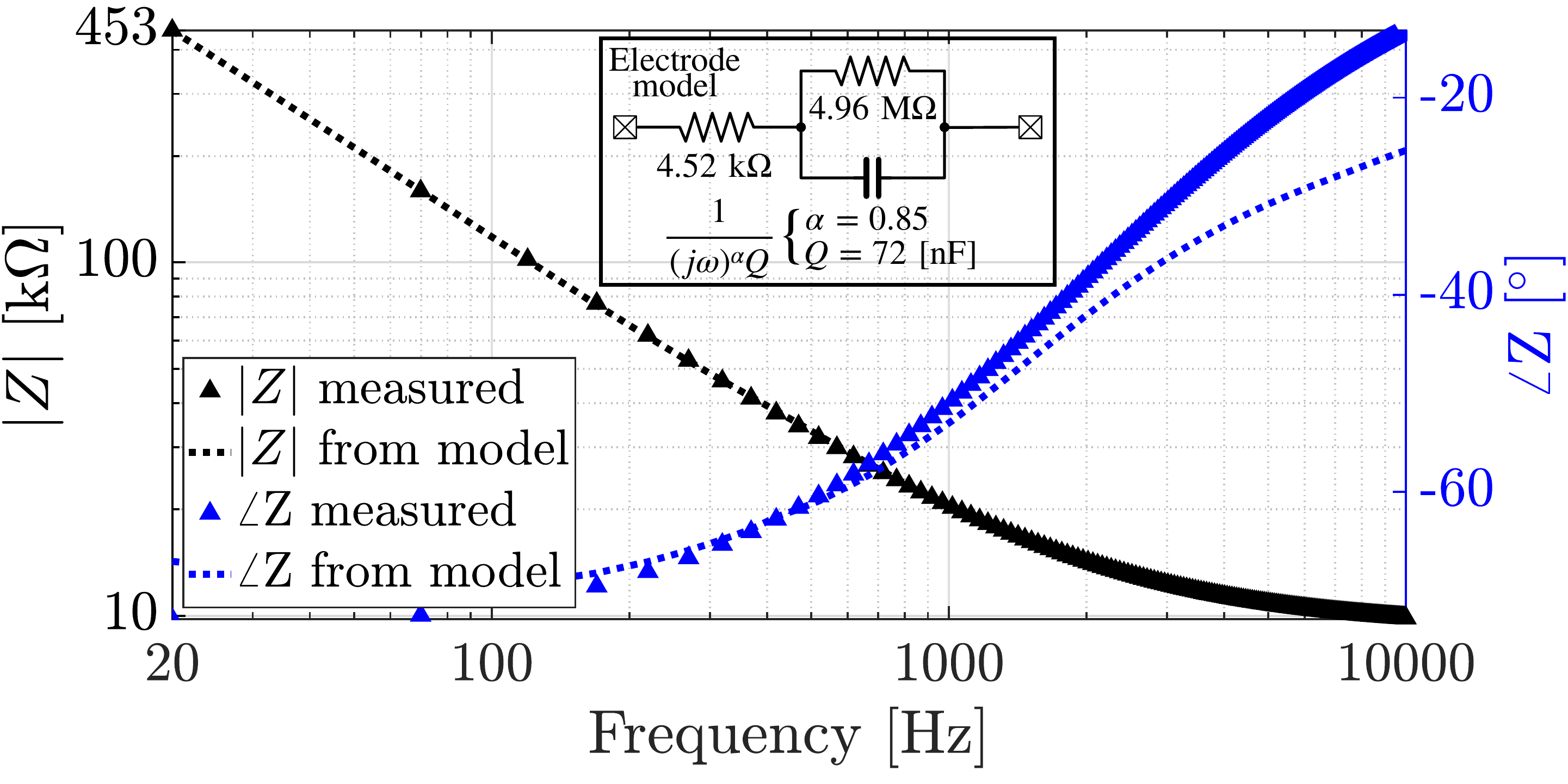}
\caption{{\color{black}Measured ENIG electrode impedance; (inset) single electrode model.}}\label{Fig:meas_electrode}
\end{figure}
\begin{figure}[t]
\centering
\includegraphics[width=1\linewidth]{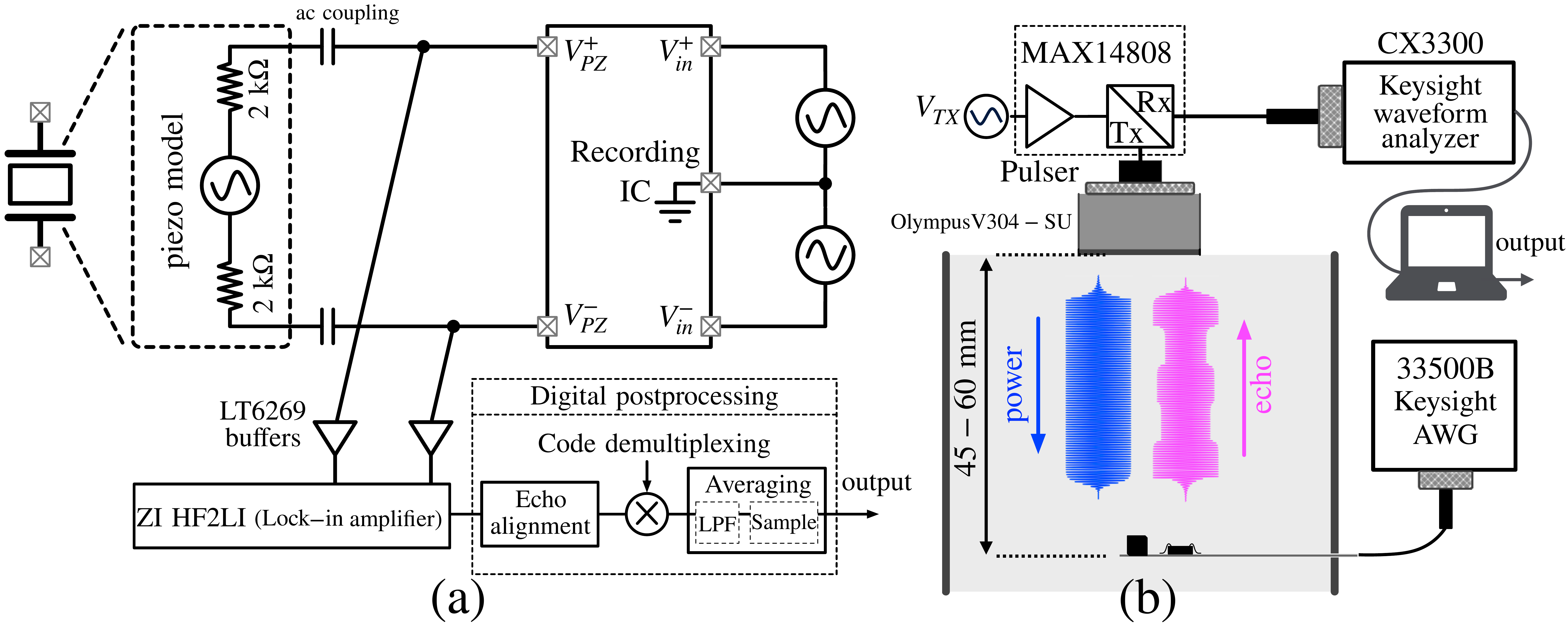}
\caption{{\color{black}(a) benchtop measurement setup, (b) single-mote \emph{in vitro} measurement setup} }\label{Fig:meas_setups}
\end{figure}

The LNA drives a linear transconductance stage for converting the acquired signal to current. The $G_m$-cell therefore requires a $\pm$160 mV input voltage range with better than 0.5\% nonlinearity. {\color{black}This is achieved by forcing the input voltage across a PFET device biased in the triode region with a differential super source follower (Fig. \ref{Fig:linera_gm})}. A PFET device is used instead of a resistor to save area without sacrificing linearity. A $\Delta$ incremental increase of $V_{in}$ results in both $|V_{DS}|$ and $|V_{GS}|$ of $M_M$ to increase by respectively $\Delta$ and $\Delta/2$, which gives rise to $I_{M_M}\propto \Delta$ as long as $M_M$ is in triode region which is ensured by designing $M_M$ as a long-channel low-Vt device. The current generated through $M_M$ creates a pair of differential current signals passing through $M_1$ and $M_2$ that is converted to single-ended in the last stage of the $G_m$-cell. The $G_m$ stage has a nominal transconductance of 120 $\mu$S. The 1 V supply powers this stage, except for the final current mirror, which is connected directly to the rectifier output and provides the signal for uplink data modulation. 2.5 V devices are used in the final mirror stage due to higher rectifier voltages. 

\subsection{Linear echo modulator}

The linear relationship between $\Gamma$ and $V_3$, expressed in (\ref{eq:gamma_approx_3}), reveals that linear amplitude modulation of the echo is possible by linearly modulating the amplitude of the piezo voltage ($V_{PZ}=V_{PZ}^+-V_{PZ}^-$ in Fig. \ref{Fig:rectifier_nonlinearity}). At resonance, the piezo is modeled by an AC voltage source ($V_s$) and the series resistance of the piezo ($R_s$). Consider the conceptual circuit diagram shown in Fig. \ref{Fig:rectifier_nonlinearity}(a) where it is assumed the signal modulating the echo amplitude is available in the current domain, $I_m$. To modulate the amplitude of the piezo voltage, $I_m$ is upmodulated by a current mixer whose switching phase is synchronous to the piezo voltage. This results in the peaks and valleys of the piezo voltage {\color{black}dropping} by $I_m R_s$. A similar AM modulation technique holds true for reflective antenna systems.

In this work, the synchronized up-conversion current mixer is implemented with minimal hardware overhead by reusing the active rectifier. The circuit diagram of the active rectifier is shown in Fig. \ref{Fig:rectifier_nonlinearity}(b) where high-frequency common-gate RF amplifiers are used as comparators. Since the rectifiers are inherently nonlinear, rectifier-induced nonlinearity should be taken into account. It can be shown that the relationship between the DC voltage at the output of the rectifier $V_{rect}$ and the load current $I_m$ is given by 
 \begin{IEEEeqnarray}{rclll}
I_m &\propto & \frac{V_s}{R_{S}}(1-\frac{V_{rect}}{V_s})(1-\frac{2}{\pi}\mathrm{asin}(\frac{V_{rect}}{V_s})),
\label{eq:rectifier}
\end{IEEEeqnarray}
where, $V_s$ is the peak value of the piezo open-circuit voltage. Equation (\ref{eq:rectifier}) is plotted in Fig. \ref{Fig:rectifier_nonlinearity}(c) where the nonlinearity between $V_{rect}$ (normalized to $V_s$) and $I_m$ (normalized to $V_s/R_s$) is shown for 10\% modulation depth. {\color{black}The} voltage across the piezo ($V_{PZ}$) is the upmodulated version of $V_{rect}$, therefore $V_{rect}$ and $V_{PZ}$ voltages are equivalent in (\ref{eq:rectifier}). For 10\% modulation depth the maximum nonlinearity between $V_{PZ}$ (and ultimately $\Gamma$) and $I_m$ is less than 0.5\%. {\color{black} Since rectifier nonlinearity was shown to be minimal, a single-ended Gm-cell was connected to the output of the rectifier, mitigating the need for the Gm-cell to sink current from both terminals of the piezo to maintain full conduction angle if connected to the input of the rectifier.}

\begin{figure*}[h]
\centering
\includegraphics[width=\textwidth]{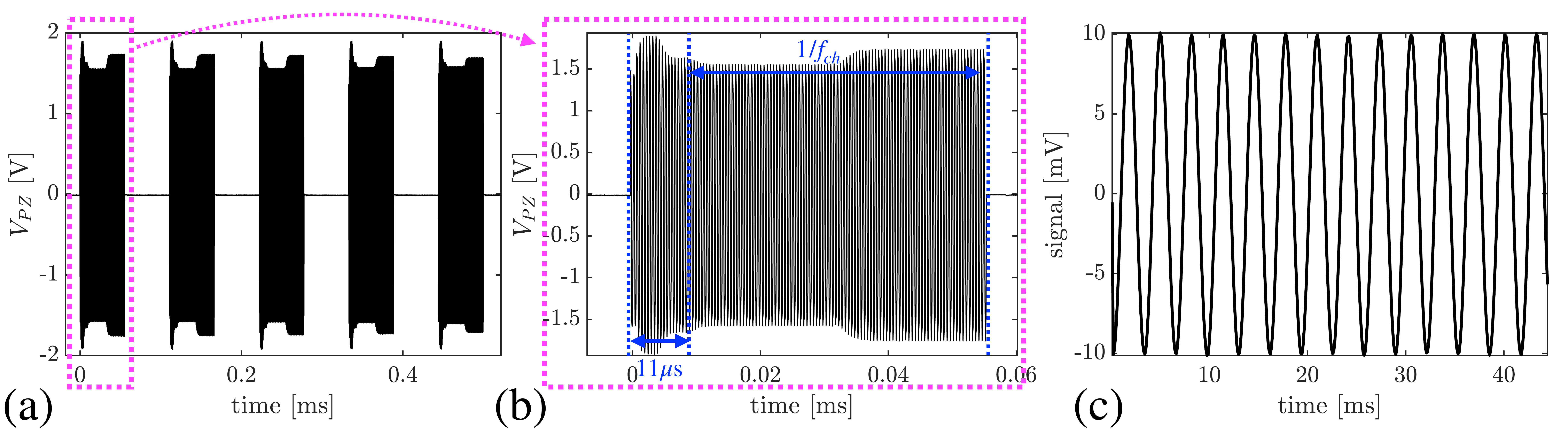}
\caption{Benchtop measurement result showing amplitude modulation of the input voltage of the rectifier for (a) five consecutive sample pulses for a 313 Hz, 20 $\mathrm{mV_{PP}}$ input signal (b) first interrogation event where startup time and the subcarrier signal are visible (c) the demodulated input-referred signal.}\label{Fig:meas_transient_only}
\end{figure*}
\begin{figure}[h]
\centering
\includegraphics[width=1\linewidth]{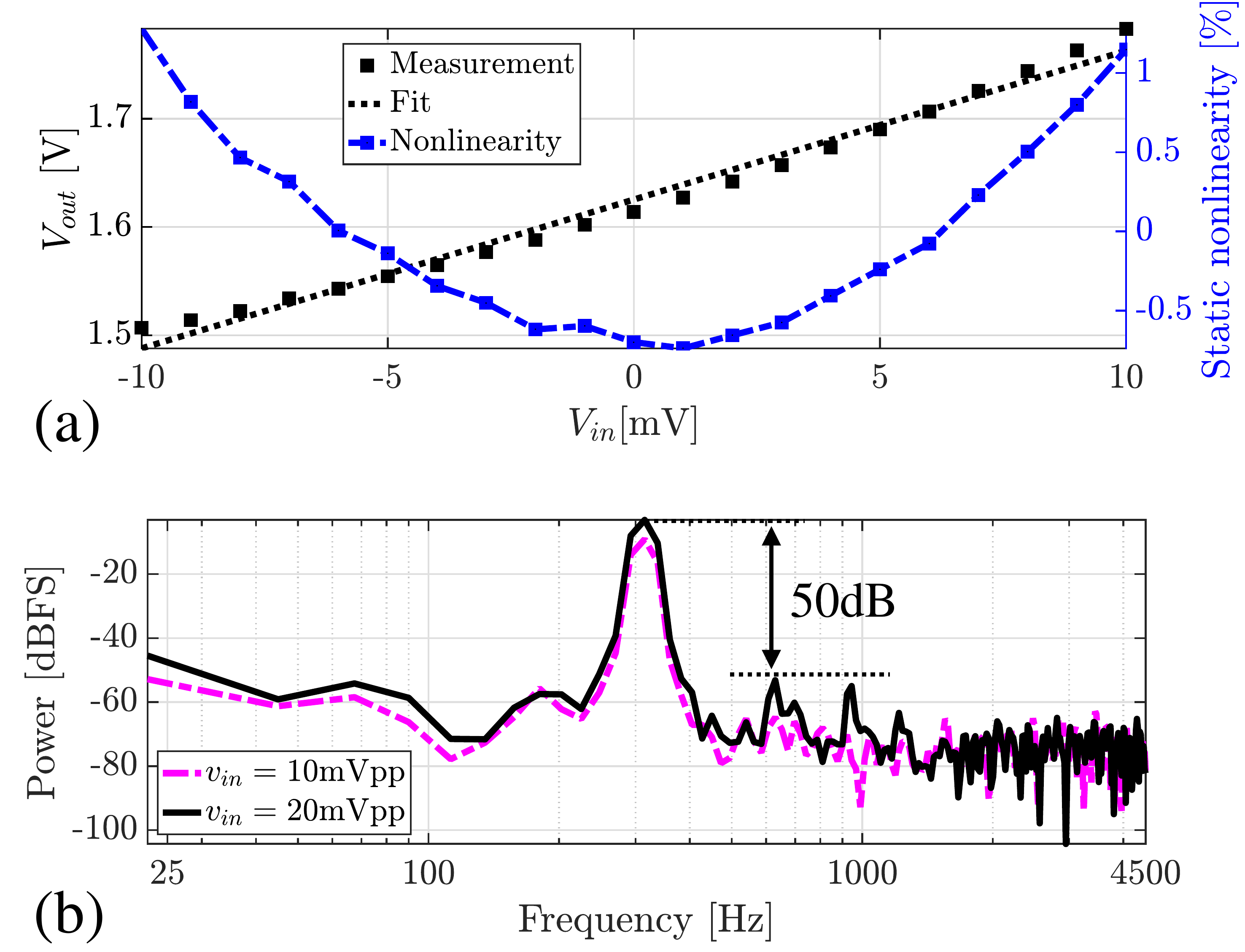}
\caption{(a) The static input-output (defined as the input voltage of the rectifier) linearity curve of the chip. (b) power spectral density of the reconstructed signal shown in Fig. \ref{Fig:meas_transient_only}.}\label{Fig:meas_linearity}
\end{figure}
\begin{figure}[h]
\centering
\includegraphics[width=1\linewidth]{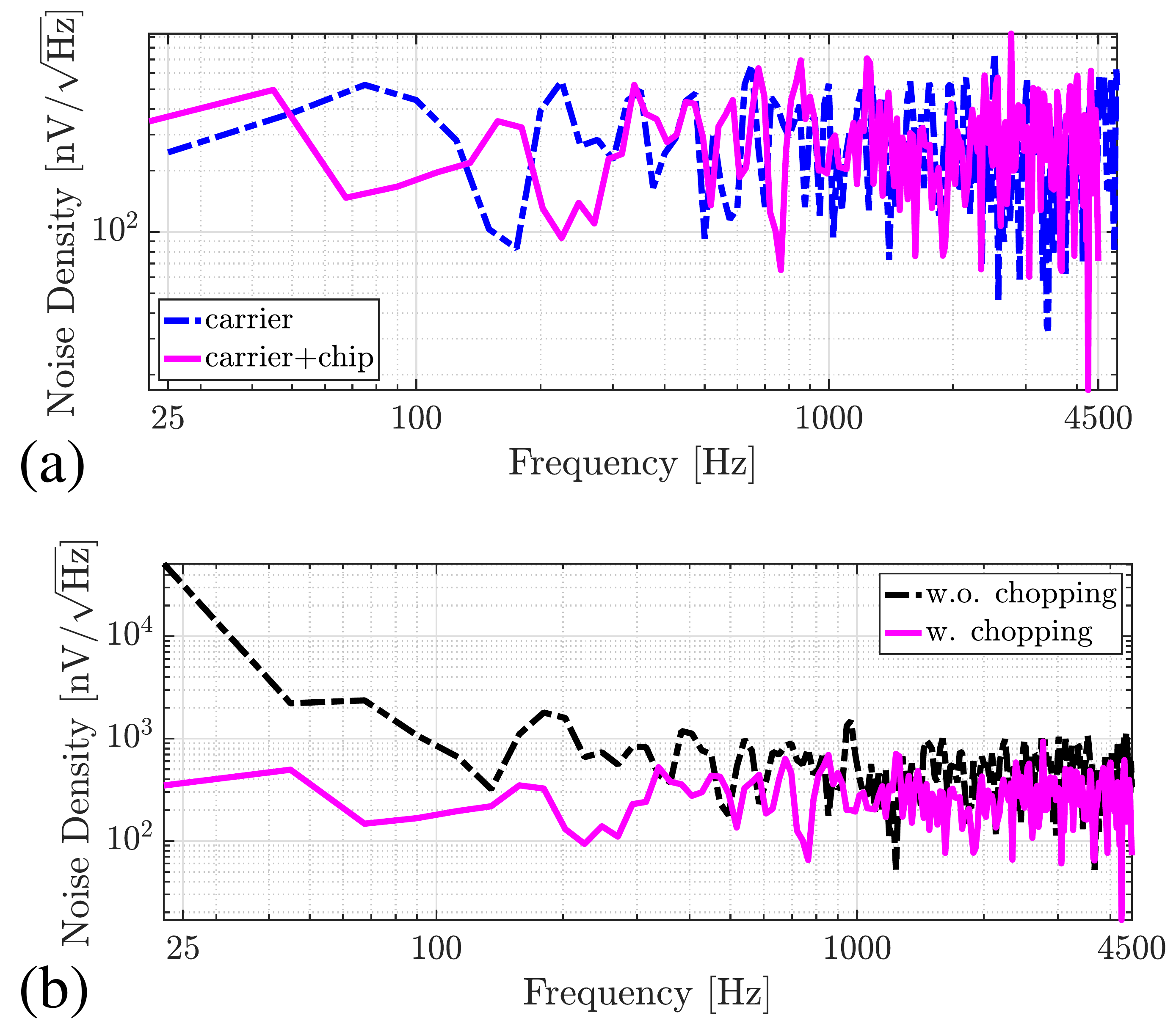}
\caption{Noise measurement. (a) the noise spectral density of the carrier alone and that obtained from the chip: noise is dominated by the carrier noise. (b) effect of chopping in reducing the low-frequency noise contents of the main carrier.}\label{Fig:meas_noise}
\end{figure}

\section{Measurement Results} \label{sec:measurement}
{\color{black}}

The IC was fabricated in the TSMC 65nm LP CMOS process.  The die micrograph and the fully assembled implant are shown in Fig. \ref{Fig:chip}. The bulk piezo and the chip are bonded to a flex PCB interposer. A pair of 200x200 $\mu$m$^2$ electroless nickel immersion gold plated electrodes {\color{black}(ENIG)} are designed on the backside of the PCB with 2 mm spacing. {\color{black}The measured impedance of the ENIG electrodes submerged in saline and a model for a single electrode are shown in Fig. \ref{Fig:meas_electrode}}.  The implant is encapsulated with $\sim$10 $\mu$m of Parylene-C \cite{loeb1977parylene}. The total area of the chip, including test pads and on-chip decoupling capacitors, is 0.25 mm$^2$. The minimum voltage amplitude required at $V_{PZ}$ {\color{black}was} measured to be 1.25 V. The circuit power dissipation after rectification {\color{black}was} $\sim$30 $\mu$W, and the total power consumption including the efficiency of the power management circuits {\color{black}were} measured to be 37.7 $\mu$W during normal operation with a 50\% duty cycle. The breakdown of the power consumption is shown in Fig. \ref{Fig:chip}(d).

{\color{black}The chip characterization setup and measurement results are shown in Fig. \ref{Fig:meas_setups}-\ref{Fig:meas_noise}. The setup includes a piezo model, an AC-coupled voltage source in series with a 4 k$\Omega$ resistor, connected to the piezo terminals of the chip. The output was measured using a fully differential lock-in amplifier for main-carrier demodulation.   Subsequent signal processing steps, e.g. subcarrier demodulation, were performed on a PC.} The output transient response of the chip, measured at the piezo voltage terminals, is shown in Fig. \ref{Fig:meas_transient_only}(a) in response to a 20 $\mathrm{mV_{PP}}$ input sine wave for five consecutive interrogation events. The first interrogation event is shown in Fig. \ref{Fig:meas_transient_only}(b) where 11 $\mu$s of power-on/startup time and the 27.5 kHz subcarrier signal are observable. The demodulated and reconstructed input signal for the same measurement is shown in Fig. \ref{Fig:meas_transient_only}(c). Static and dynamic nonlinearity measurement results are shown in Fig. \ref{Fig:meas_linearity}. An end-to-end voltage gain ($\Delta V_{PZ}/v_{in}$) of 23 dB with maximum static non-linearity error of 1.2\% {\color{black}was} measured. The power spectrum of the reconstructed {\color{black}313} Hz, 20 $\mathrm{mV_{PP}}$ sine wave is shown in Fig. \ref{Fig:meas_linearity}(b) and {\color{black}achieved} an SFDR of 50 dB and a THD of -44 dB. No harmonic tones are visible for a 10 $\mathrm{mV_{PP}}$ input signal.

The noise measurement results are summarized in Fig. \ref{Fig:meas_noise} for an interrogation (sample) frequency of 10 kHz. There are two major contributors to the total input referred noise density: the AM noise of the carrier, and the noise contributed by the recording circuits. The total input referred noise spectral density {\color{black}was} measured to be 328 $\mathrm{nV/\sqrt{Hz}}$. This is mainly dominated by the carrier noise, measured at 319 $\mathrm{nV/\sqrt{Hz}}$ in the absence of the chip (Fig. \ref{Fig:meas_noise}(a)). Both the curves in Fig. \ref{Fig:meas_noise}(a) are derived by down-chopping and averaging, which partially removes the 1/f noise of the carrier. Assuming the noise of the carrier and that of the chip are additive, the input referred noise of the chip alone can be estimated ~76 $\mathrm{nV/\sqrt{Hz}}$, or 5.37 $\mu\mathrm{V_{RMS}}$ in a 5 kHz bandwidth. The effect of chopping in bypassing the low-frequency noise contents of the main carrier is demonstrated in Fig.  \ref{Fig:meas_noise}(b), where 1/f noise is clearly visible in the spectrum when chopping is disabled.

\begin{figure*}[t]
\centering
\includegraphics[width=1\linewidth]{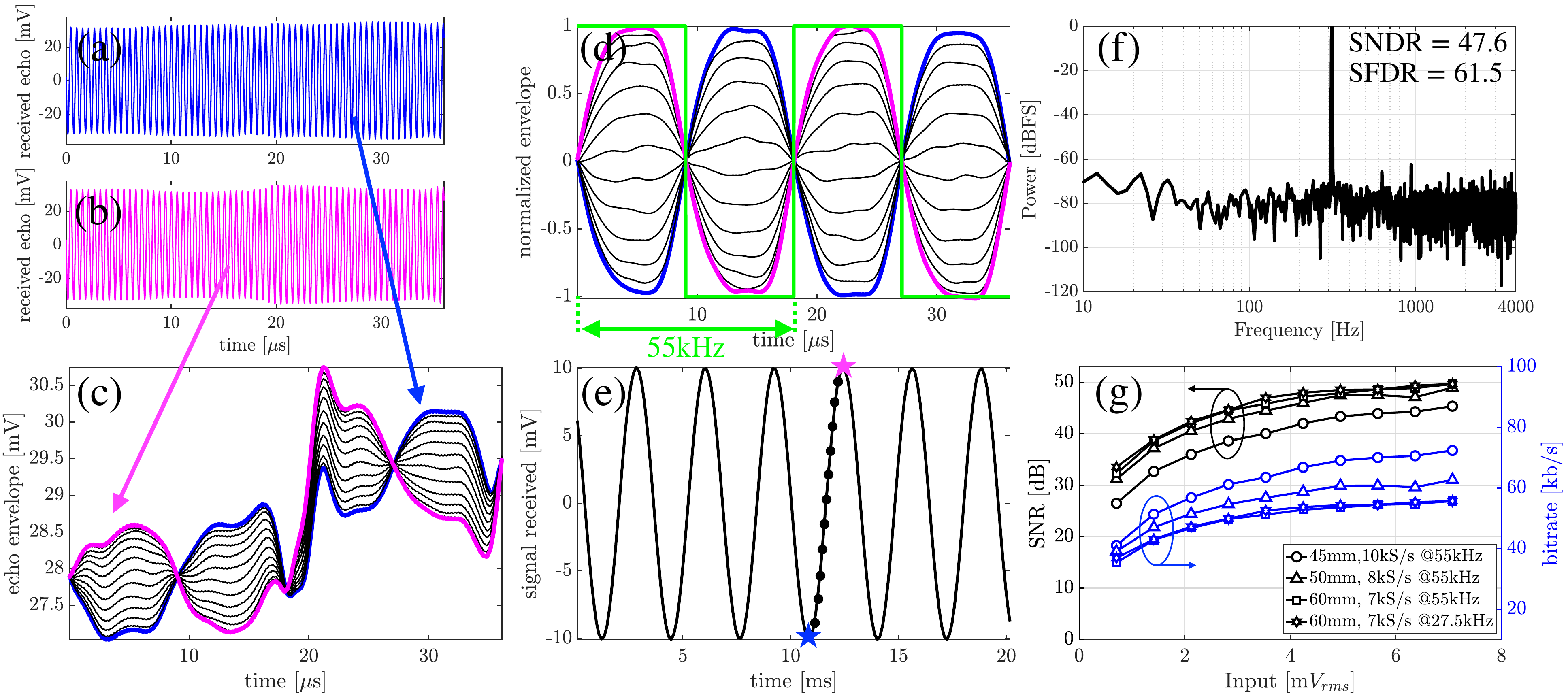}
\caption{{\color{black}Single-mote \emph{in vitro} measurement results. Implant is interrogated at 8 kS/s at 5 cm of depth (a-b) two received echo signals each translating into a sample shown in (e), (c) their corresponding AM demodulation (d) common-mode rejection of received echoes (e) reconstructed 313 Hz signal (f) SNR of received signal (noise dominated by carrier noise) and (g) measured SNR and equivalent uplink data rate vs. input range of the implant. }}\label{Fig:meas_single_invitro}
\end{figure*}
\begin{figure}[t]
\centering
\includegraphics[width=1\linewidth]{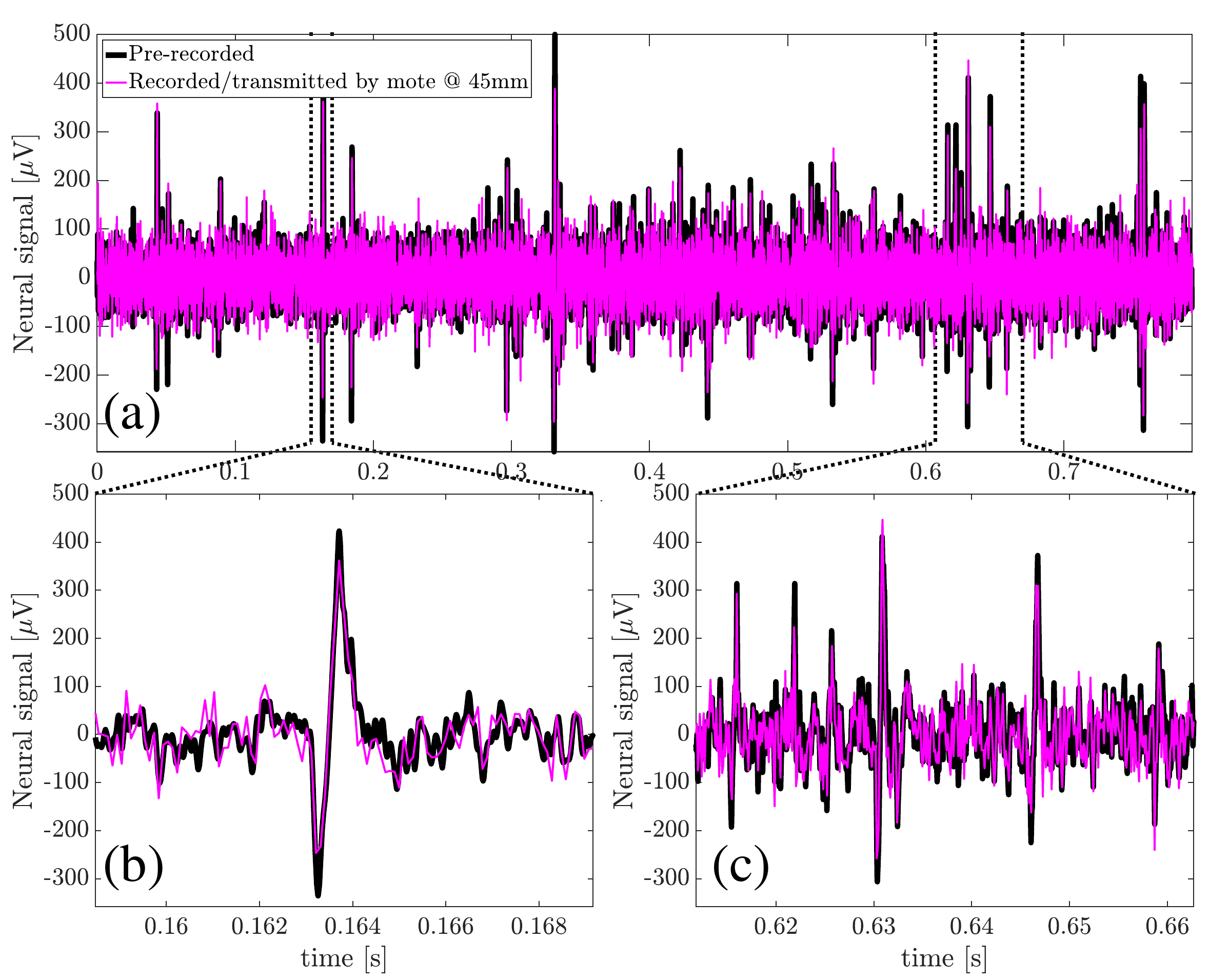}
\caption{{\color{black}(a) 800 ms stream of pre-recorded neural signal recorded (at 10 kS/s) and wirelessly transmitted by the mote at 45 mm of depth. Comparison of reconstructed data between (b) single and (c) multiple action potential events.}}\label{Fig:meas_neural}
\end{figure}

{\color{black}The single-mote} \emph{in vitro} measurement setup is shown in Fig. \ref{Fig:meas_setups}(b), where a single assembled mote is suspended at a distance of 45--60 mm away from a single-element external transducer in {\color{black}oil (with $\sim$0.5 dB/cm attenuation at 2MHz). The implant was interrogated at 8 kS/s.} The main carrier frequency {\color{black}was} set to the resonant frequency of the implant piezo at 1.78 MHz. A subcarrier frequency of 55 kHz was generated on chip. Two received sample echoes that form a peak and a valley of {\color{black}a 313 Hz}, 20 $\mathrm{mV_{PP}}$  signal are shown in Fig. \ref{Fig:meas_single_invitro} along with their demodulated signals. The reconstructed received signal and its spectrum are also shown in Fig. \ref{Fig:meas_single_invitro}. {\color{black}Although the noise of the carrier (generated by the external ultrasound pulser) dominates the overall noise of the link, for the 20 $\mathrm{mV_{PP}}$ input range of the implant, 47.96 dB of SNR is measured. Were the carrier noise absent, the SNR would improve by $\sim$10-12 dB. Fig. \ref{Fig:meas_single_invitro}(g) summarizes the measured SNR and equivalent uplink data rate measured at multiple other possible configurations with varied depth, interrogation frequency and subcarrier frequency using the same setup introduced earlier.}

{\color{black}To make sure an even number of subcarrier cycles (e.g. 2 in Fig. \ref{Fig:meas_single_invitro}(d)) is used for demodulation; first, echo duration is chosen to be sufficiently long. For instance, \textgreater 29 $\mu$s and \textgreater 47 $\mu$s echoes are needed for 2 and 4 cycles of a 55 kHz sub carrier respectively. Since the startup time and consequently the start of the echo modulation period of the chip is consistent from sample to sample, and since the period of the sub carrier is known and referenced to the main carrier frequency (e.g. 1.78 MHz$\div$32 = 55 kHz), the subcarrier signal can be determined at the interrogator receiver for demodulation. For pulses longer than the ones mentioned above (29 $\mu$s and 47 $\mu$s), the received echoes are truncated to 29 $\mu$s and 47 $\mu$s such that only an even number of cycles are used for demodulation. 

Further \emph{in vitro} verification of the mote was performed where an 800 ms stream of pre-recorded neural signal (acquired by Plexon multichannel acquisition processor) from an awake-behaving rat (Long-Evans) motor cortex was fed to the chip and wirelessly transmitted to the external interrogator. The mote was placed at the depth of 45 mm in a tissue phantom (with $\sim$0.5 dB/cm attenuation at 2 MHz) and interrogated at 10 kS/s. Fig. 20 shows the comparison between the reference pre-recorded neural signal with the signal recorded and wirelessly transmitted by the mote.}

\begin{figure*}[t]
\centering
\includegraphics[width=1\linewidth]{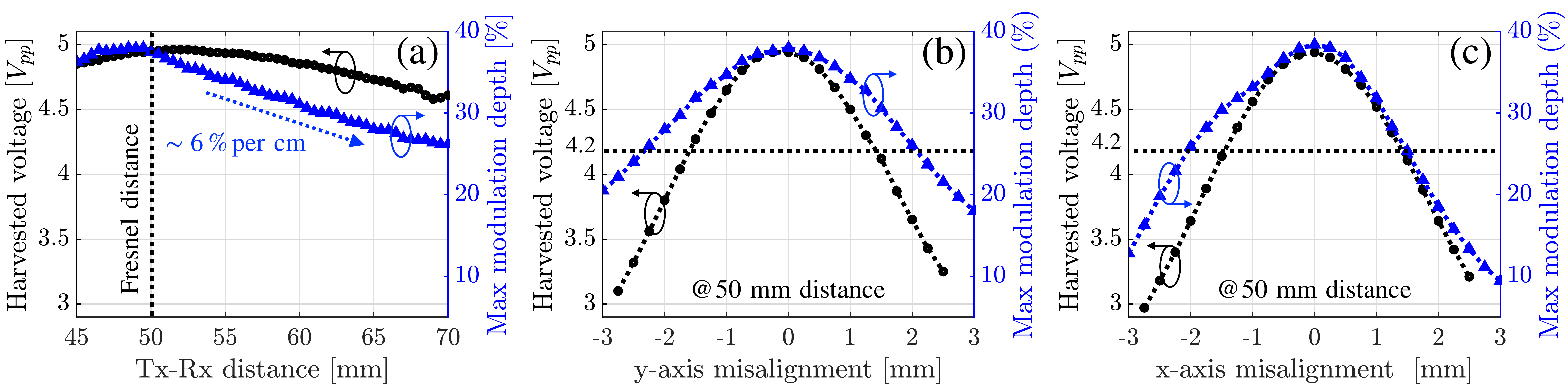}
\caption{{\color{black}Measured piezo-interrogator relative misalignment characterization in (a) vertical (b) horizontal y-axis (c) horizontal x-axis dimensions}}\label{Fig:meas_misalignment}
\end{figure*}
\begin{figure*}[h]
\centering
\includegraphics[width=1\linewidth]{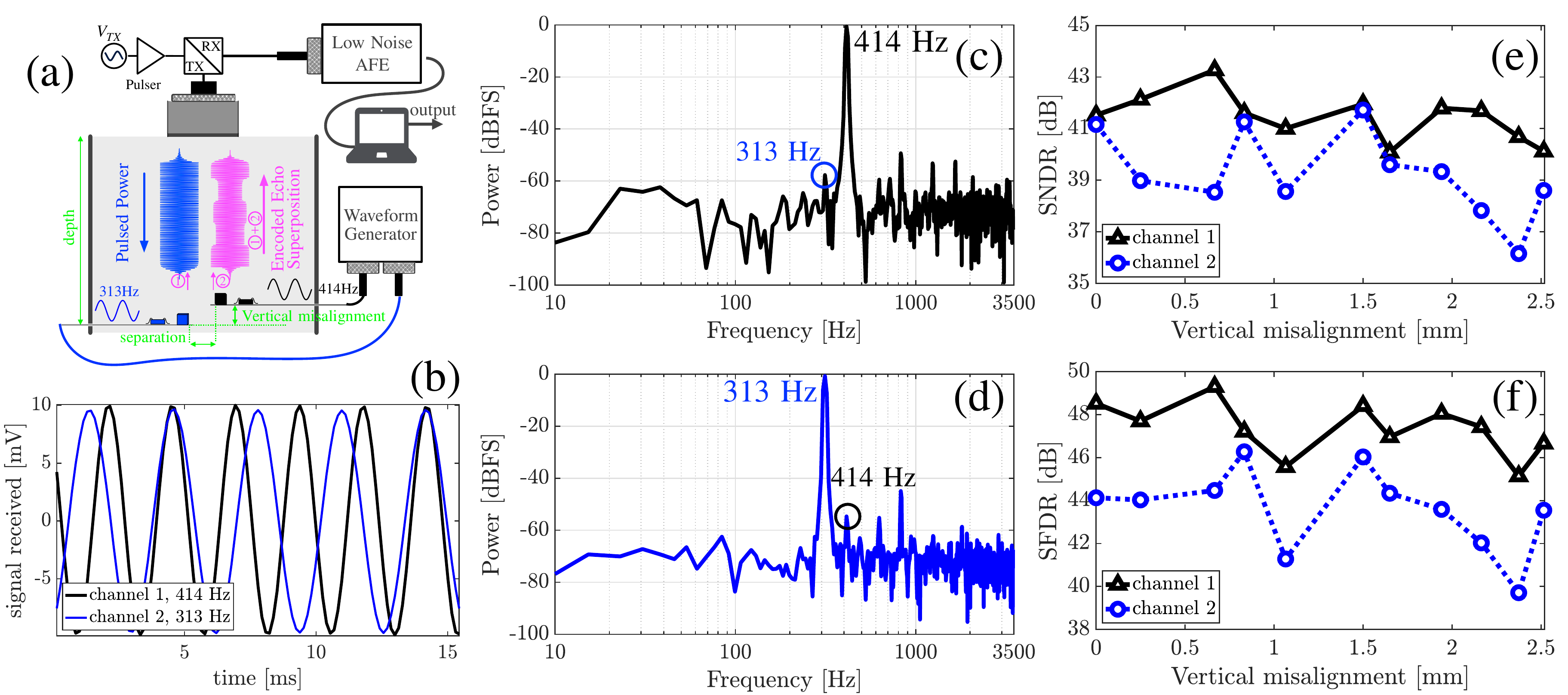}
\caption{{\color{black}Simultaneous power up and data transmission of two implants at depth of 50 mm and separation of 2 mm. (a): in vitro measurement setup. (b): reconstructed signals at the external interrogator. (c-d) spectra of the reconstructed signals (e-f) measured SNDR and SFDR of each channel versus vertical misalignment between the motes}}\label{Fig:meas_multi_invitro}
\end{figure*}
\begin{figure}[!h]
\centering
\includegraphics[width=1\linewidth]{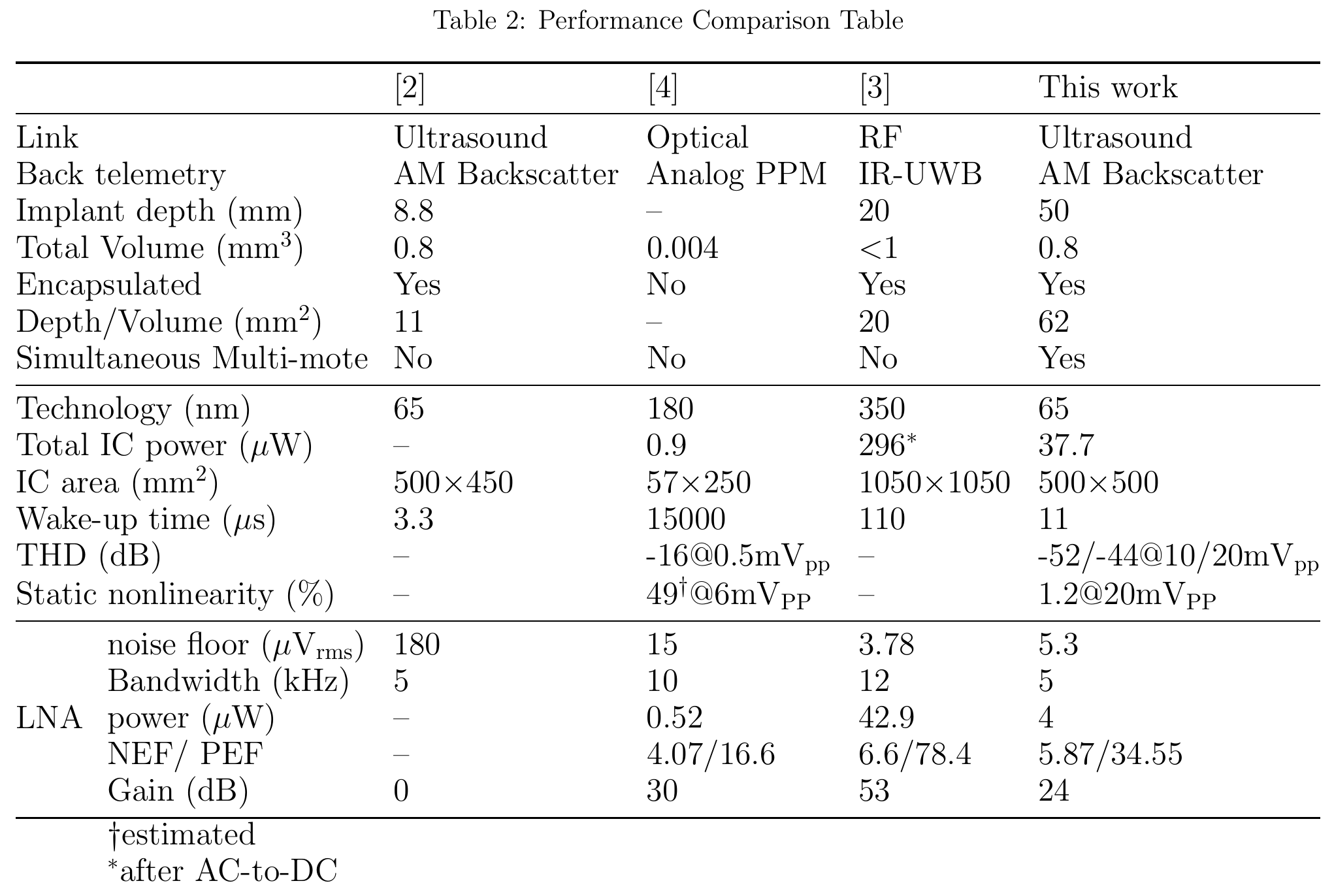}
\end{figure}

{\color{black}The effects of misalignment on the operating range of the device are characterized in Fig. \ref{Fig:meas_misalignment}. A set of measurement results reporting the harvested piezo voltage and the maximum modulation depth of the echo at various relative locations of the external transducer and the implant piezo. The measurement medium was oil with $\sim$0.5 dB/cm attenuation at 2 MHz. An unfocused 0.5$''$ diameter external transducer was driven at 1.78 MHz by a $\pm$15 V pulser. The maximum modulation depth is defined based on the received echo amplitude at two extreme piezo terminations, open- and short-circuited piezos. Fig. \ref{Fig:meas_misalignment}(a) illustrates that the harvested power is maximized at the Fresnel distance of the external transducer ($\sim$ 52 mm), and that the optimal operation depth of the implants is $\sim$50 mm where the harvested voltage and the modulation depth are concurrently large. Moreover, at 70 mm of depth, acceptable harvested voltages (\textgreater 4.5 Vpp) and maximum modulation depth (\textgreater20 \%) were measured. It is also observed that beyond the Fresnel distance, the modulation depth steadily decays at a rate of 6\% per cm, allowing a large range of viable implant depths for the mote. Fig. \ref{Fig:meas_misalignment}(b) and (c) demonstrate similar measurement results for horizontal x- and y-axis misalignment between the implant piezo and the external transducer. It can be observed that the effect of horizontal misalignment is symmetric with the respect to the line of sight (LoS); a slight asymmetry in Fig. \ref{Fig:meas_misalignment}(c) is due to the setup, which includes a rod holding the piezo along the x-axis. Since the chip is fully operational at harvested voltages greater than 4.2 Vpp, horizontal misalignment of up to $\pm$1.75 mm is acceptable at a cost of a negligible drop in modulation depth. The $\pm$1.75 mm misalignment margin is directly proportional to the aperture of the external transducer, therefore the margin can be doubled by using a 1$''$ external transducer.}

{\color{black} Fig. \ref{Fig:meas_multi_invitro} demonstrates an \emph{in vitro} measurement setup where two implants at a depth of 50 mm with a 2 mm separation were synchronously powered up by a single 0.5$''$  unfocused external transducer. The setup environment limits the depth in this dual-mote measurement. The subcarrier frequencies of the implants are orthogonal to each other (55 kHz and 27.5 kHz) to enable simultaneous uplink data transmission in this dual-mote setup. Each implant transmitted a single tone (414 Hz and 313 Hz) to the external receiver. The reconstructed tones are shown in Fig. \ref{Fig:meas_multi_invitro}(b) along with their spectra. Measured SNDR and SFDR are shown as a function of vertical misalignment between motes in Fig. \ref{Fig:meas_multi_invitro}(e) and (f), respectively. It can be observed that the uplink data transmission quality is maintained over $\pm$2 mm of vertical misalignment. A given vertical misalignment $\Delta$Z between two implants, results in the implants powering up with a delay equal to $\Delta$t $=\Delta$Z/c, where c is the propagation speed of sound in tissue. This delay, in turn, results in subcarriers becoming out of phase by $\Delta$t, which translates to inter-channel crosstalk and consequently degradation of SNDR and SFDR. 
Given the carrier frequency and the highest frequency of the orthogonal codes are 1.78 MHz and 55 kHz respectively, at depths between 35 and 70 mm, 4 implants can be simultaneously interrogated.  At depths $\geq$70 mm, up to 8 devices can be simultaneously interrogated.}

\section{Summary} \label{sec:summary}
We present a 0.8 mm$^3$ free floating implant that uses a single ultrasound link for wireless power harvesting and analog data back telemetry. The theoretical basis for a linear amplitude modulated ultrasound echo modulation technique was introduced, achieving a 20  $\mathrm{mV_{PP}}$ linear range of the implant. A comparison with recently published fully integrated free-floating sub-mm$^3$ neural recording implants is shown in Table 2. This work advances the noise performance of \cite{seo2016wireless} by 34x without sacrificing the implant volume. Compared to prior art \cite{seo2016wireless, yeon2018towards, lee2018250mum}, this work achieves the lowest nonlinearity at the highest input range, achieves a comparable NEF to that of the state-of-the-art, and improves operating depth by \textgreater 2.5x, resulting in the highest measured depth/volume ratio by $\sim$3x. We demonstrate, for the first time to our knowledge, simultaneous power-up and communication with two free-floating motes without requiring a specialized, e.g. beamformed, external transducer.

\appendix[Piezo $\Gamma$ vs. Voltage linearity]
This section provides derivation of (\ref{eq:gamma_approx_33}) earlier introduced in Section \ref{sec:linear_echo}. 
The piezo is modeled as a thickness-mode resonating 3-port network, shown in Fig. \ref{Fig:matrix}(a), whose input-output port relationships are well described by \cite{kino1987acoustic},
\begin{IEEEeqnarray}{rclll}
	\begin{bmatrix}
    F_1 \\
    F_2 \\
    V_3 \\
   \end{bmatrix} &=& \mathbf{P}\begin{bmatrix}
    \nu_1 \\
    \nu_2 \\
    I_3 \\
   \end{bmatrix}&=& 
	   	\begin{bmatrix}
   m & n & p \\
   n & m & p \\
   p & p & r \\
   \end{bmatrix}
   	\begin{bmatrix}
    \nu_1 \\
    \nu_2 \\
    I_3 \\
   \end{bmatrix}
\end{IEEEeqnarray}
\begin{IEEEeqnarray}{rcl}
\begin{bmatrix}
    F_1 \\
    F_2 \\
    V_3 \\
   \end{bmatrix} &=&
   \begin{bmatrix}
    \frac{Z_0A}{j\mathrm{tan}(\beta l)} & \frac{Z_0A}{j\mathrm{sin}(\beta l)} & \frac{h_{33}}{j\omega} \\
    \frac{Z_0A}{j\mathrm{tan}(\beta l)} & \frac{Z_0A}{j\mathrm{sin}(\beta l)} & \frac{h_{33}}{j\omega} \\
    \frac{h_{33}}{j\omega} & \frac{h_{33}}{j\omega}& \frac{1}{j\omega C_0} \\
   \end{bmatrix}
   \begin{bmatrix}
    \nu_1 \\
    \nu_2 \\
    I_3 \\
   \end{bmatrix}
   \label{eq:first}
\end{IEEEeqnarray}
Ports 1 and 2 are acoustical, and port 3 is the electrical port of the piezo. Table \ref{Tab:param} describes the parameters used in (\ref{eq:first}). The acoustic impedance seen into port 1, while port 2 and 3 are respectively terminated by $Z_B$ and $Z_E$ (Fig. \ref{Fig:matrix}(b)) is given by
\begin{IEEEeqnarray}{rcl}
Z_1 &=& \frac{p^2(2n-2m-Z_B)+(Z_E+r)(m^2-n^2+mZ_B)}{(Z_E+r)(m+Z_B)-p^2}.
\end{IEEEeqnarray}
Therefore, $\Gamma$ defined as $\Gamma=(Z_1-Z_B)/(Z_1+Z_B)$ becomes
\begin{IEEEeqnarray}{rcl}
\Gamma &=& \frac{2p^2(n-m)+(Z_E+r)(m^2-n^2-Z_B^2)}{2p^2(n-m-Z_B)+(Z_E+r)((m+Z_B)^2-n^2)}.
\label{eq:gamma_act}
\end{IEEEeqnarray}
Rearranging terms and noting that $m^2-n^2=(Z_oA)^2 \gg Z_B^2$, \eqref{eq:gamma_act} can be simplified to 
\begin{IEEEeqnarray}{rcl}
\Gamma &\approx & \frac{Z_E+Z_{3,NL}}{Z_E+2Z_B(\frac{mr-p^2}{m^2-n^2})+Z_{3,NL}},
\label{eq:gamma_approx_1}
\end{IEEEeqnarray}
where $Z_{3,NL}$ is the impedance seen into port 3 when ports 1 and 2 are acoustically unloaded  (Fig. \ref{Fig:matrix}(d), $Z_B=0$). In fact, for nonzero acoustic termination impedance at ports 1 and 2 (Fig. \ref{Fig:matrix}(c)),
\begin{IEEEeqnarray}{rcl}
Z_3 &=& r-\frac{2p^2}{m+n+Z_B}.
\end{IEEEeqnarray}
The series resonant frequency ($f_s$) is defined as the frequency at which the impedance seen into the electrical port of an acoustically unloaded piezo is real. That is, at $f_s$, $Z_{3,NL}=0$ ($Z_{3,NL}=Z_3 (@Z_B=0)$ is purely imaginary), and (\ref{eq:gamma_approx_1}) further simplifies to
\begin{IEEEeqnarray}{rcl}
\Gamma &\approx & \frac{Z_E}{Z_E+2Z_B(\frac{mr-p^2}{m^2-n^2})}.
\label{eq:gamma_approx_2}
\end{IEEEeqnarray}
Moreover, at $f_s$, the piezo resonator is modeled by a voltage source and a series resistance, shown in Fig. \ref{Fig:analog_linearity}(b), whose value is given by
\begin{IEEEeqnarray}{rclllll}
R_{S} &=& \mathrm{Re\{Z_3\}}&=&\mathrm{Re}\{r-\frac{2p^2}{m+n+Z_B}\}&\approx& \frac{-2Z_Bp^2}{(n+m)^2} .
\label{eq:R_piezo}
\end{IEEEeqnarray}
Dividing the second term in the denominator of (\ref{eq:gamma_approx_2}) by (\ref{eq:R_piezo}) results in $(1-mr/p^2 )((m+n)/(m-n))$ which equals 1 at $f_s$, because
\begin{IEEEeqnarray}{rcl}
\frac{m+n}{m-n} &=& \mathrm{cot}^2(\frac{\beta l}{2})
\end{IEEEeqnarray}
and
\begin{IEEEeqnarray}{rcl}
1-\frac{mr}{p^2} &=& \frac{\beta l}{k^2\mathrm{tan}(\beta l)}.
\label{eq:second_RS}
\end{IEEEeqnarray}
Given at $f_s$ \cite{kino1987acoustic},
\begin{IEEEeqnarray}{rcl}
\frac{\mathrm{tan}(\beta l/2)}{\beta l/2}&=&\frac{1}{k_T^2}=\frac{1+k^2}{k^2},
\end{IEEEeqnarray}
(\ref{eq:second_RS}) can be further simplified to 
\begin{IEEEeqnarray}{rcl}
1-\frac{mr}{p^2} &=& \mathrm{tan^2}(\frac{\beta l}{2})
\end{IEEEeqnarray}
Therefore, the second term in the denominator of \eqref{eq:gamma_approx_2} and $R_{S}$ given by \eqref{eq:R_piezo} are equal. 
That is, 
\begin{IEEEeqnarray}{rclllll}
R_{piezo,S} &=& \frac{-2Z_Bp^2}{(n+m)^2}&=&2Z_B(\frac{mr-p^2}{m^2-n^2}).
\end{IEEEeqnarray}
Therefore \eqref{eq:gamma_approx_2} can be rewritten as
 \begin{IEEEeqnarray}{rclll}
\Gamma &\approx & \frac{Z_E}{Z_E+R_{S}}&\propto &V_3.
\label{eq:gamma_approx_3}
\end{IEEEeqnarray}

%



\section*{Acknowledgment}
The authors thank DARPA BTO, the sponsors of Berkeley Wireless Research Center and TSMC for chip fabrication. Thanks to Ka Yiu Lee and Burak Eminoglu for technical discussion.
%
%

%


\bibliographystyle{IEEEtran}
\bibliography{IEEEabrv,./Myref}

\begin{thebibliography}{10}
\providecommand{\url}[1]{#1}
\csname url@samestyle\endcsname
\providecommand{\newblock}{\relax}
\providecommand{\bibinfo}[2]{#2}
\providecommand{\BIBentrySTDinterwordspacing}{\spaceskip=0pt\relax}
\providecommand{\BIBentryALTinterwordstretchfactor}{4}
\providecommand{\BIBentryALTinterwordspacing}{\spaceskip=\fontdimen2\font plus
\BIBentryALTinterwordstretchfactor\fontdimen3\font minus
  \fontdimen4\font\relax}
\providecommand{\BIBforeignlanguage}[2]{{%
\expandafter\ifx\csname l@#1\endcsname\relax
\typeout{** WARNING: IEEEtran.bst: No hyphenation pattern has been}%
\typeout{** loaded for the language `#1'. Using the pattern for}%
\typeout{** the default language instead.}%
\else
\language=\csname l@#1\endcsname
\fi
#2}}
\providecommand{\BIBdecl}{\relax}
\BIBdecl

\bibitem{maharbiz2017reliable}
M.~M. Maharbiz, R.~Muller, E.~Alon, J.~M. Rabaey, and J.~M. Carmena, ``Reliable
  next-generation cortical interfaces for chronic brain--machine interfaces and
  neuroscience,'' \emph{Proceedings of the IEEE}, vol. 105, no.~1, pp. 73--82,
  2017.

\bibitem{seo2016wireless}
D.~Seo, R.~M. Neely, K.~Shen, U.~Singhal, E.~Alon, J.~M. Rabaey, J.~M. Carmena,
  and M.~M. Maharbiz, ``Wireless recording in the peripheral nervous system
  with ultrasonic neural dust,'' \emph{Neuron}, vol.~91, no.~3, pp. 529--539,
  2016.

\bibitem{yeon2018towards}
P.~Yeon, M.~S. Bakir, and M.~Ghovanloo, ``Towards a 1.1 mm 2 free-floating
  wireless implantable neural recording soc,'' in \emph{2018 IEEE Custom
  Integrated Circuits Conference (CICC)}.\hskip 1em plus 0.5em minus
  0.4em\relax IEEE, 2018, pp. 1--4.

\bibitem{lee2018250mum}
S.~Lee, A.~J. Cortese, A.~P. Gandhi, E.~R. Agger, P.~L. McEuen, and A.~C.
  Molnar, ``A 250$\mu$m$\times$ 57$\mu$m microscale opto-electronically
  transduced electrodes (motes) for neural recording,'' \emph{IEEE transactions
  on biomedical circuits and systems}, vol.~12, no.~6, pp. 1256--1266, 2018.

\bibitem{ersen2015chronic}
A.~Ersen, S.~Elkabes, D.~S. Freedman, and M.~Sahin, ``Chronic tissue response
  to untethered microelectrode implants in the rat brain and spinal cord,''
  \emph{Journal of neural engineering}, vol.~12, no.~1, p. 016019, 2015.

\bibitem{pothof2017comparison}
F.~Pothof, L.~Chauvi{\`e}re, T.~Holzhammer, A.~Aarts, O.~Paul, W.~Singer, and
  P.~Ruther, ``Comparison of the in-vivo neural recording quality of floating
  and skull-fixed silicon probes,'' in \emph{2017 8th International IEEE/EMBS
  Conference on Neural Engineering (NER)}.\hskip 1em plus 0.5em minus
  0.4em\relax IEEE, 2017, pp. 158--161.

\bibitem{muller2015minimally}
R.~Muller, H.-P. Le, W.~Li, P.~Ledochowitsch, S.~Gambini, T.~Bjorninen,
  A.~Koralek, J.~M. Carmena, M.~M. Maharbiz, E.~Alon \emph{et~al.}, ``A
  minimally invasive 64-channel wireless $\mu$ecog implant,'' \emph{IEEE
  Journal of Solid-State Circuits}, vol.~50, no.~1, pp. 344--359, 2015.

\bibitem{ha2017silicon}
S.~Ha, A.~Akinin, J.~Park, C.~Kim, H.~Wang, C.~Maier, P.~P. Mercier, and
  G.~Cauwenberghs, ``Silicon-integrated high-density electrocortical
  interfaces,'' \emph{Proceedings of the IEEE}, vol. 105, no.~1, pp. 11--33,
  2017.

\bibitem{lee2019implantable}
J.~Lee, E.~Mok, J.~Huang, L.~Cui, A.-H. Lee, V.~Leung, P.~Mercier,
  S.~Shellhammer, L.~Larson, P.~Asbeck \emph{et~al.}, ``An implantable wireless
  network of distributed microscale sensors for neural applications,'' in
  \emph{2019 9th International IEEE/EMBS Conference on Neural Engineering
  (NER)}.\hskip 1em plus 0.5em minus 0.4em\relax IEEE, 2019, pp. 871--874.

\bibitem{khalifa2018microbead}
A.~Khalifa, Y.~Karimi, Q.~Wang, S.~Garikapati, W.~Montlouis,
  M.~Stana{\'c}evi{\'c}, N.~Thakor, and R.~Etienne-Cummings, ``The microbead: A
  highly miniaturized wirelessly powered implantable neural stimulating
  system,'' \emph{IEEE transactions on biomedical circuits and systems},
  vol.~12, no.~3, pp. 521--531, 2018.

\bibitem{hammer2018cervical}
N.~Hammer, S.~L{\"o}ffler, Y.~O. Cakmak, B.~Ondruschka, U.~Planitzer,
  M.~Schultz, D.~Winkler, and D.~Weise, ``Cervical vagus nerve morphometry and
  vascularity in the context of nerve stimulation-a cadaveric study,''
  \emph{Scientific reports}, vol.~8, no.~1, p. 7997, 2018.

\bibitem{wang2017closed}
M.~L. Wang, T.~C. Chang, T.~Teisberg, M.~J. Weber, J.~Charthad, and
  A.~Arbabian, ``Closed-loop ultrasonic power and communication with multiple
  miniaturized active implantable medical devices,'' in \emph{2017 IEEE
  International Ultrasonics Symposium (IUS)}.\hskip 1em plus 0.5em minus
  0.4em\relax IEEE, 2017, pp. 1--4.

\bibitem{ghanbari201917}
M.~M. Ghanbari, D.~K. Piech, K.~Shen, S.~F. Alamouti, C.~Yalcin, B.~C. Johnson,
  J.~M. Carmena, M.~M. Maharbiz, and R.~Muller, ``17.5 a 0.8 mm 3 ultrasonic
  implantable wireless neural recording system with linear am backscattering,''
  in \emph{2019 IEEE International Solid-State Circuits
  Conference-(ISSCC)}.\hskip 1em plus 0.5em minus 0.4em\relax IEEE, 2019, pp.
  284--286.

\bibitem{fry1978acoustical}
F.~Fry and J.~Barger, ``Acoustical properties of the human skull,'' \emph{The
  Journal of the Acoustical Society of America}, vol.~63, no.~5, pp.
  1576--1590, 1978.

\bibitem{chang201727}
T.~C. Chang, M.~L. Wang, J.~Charthad, M.~J. Weber, and A.~Arbabian, ``27.7 a
  30.5 mm 3 fully packaged implantable device with duplex ultrasonic data and
  power links achieving 95kb/s with< 10- 4 ber at 8.5 cm depth,'' in \emph{2017
  IEEE International Solid-State Circuits Conference (ISSCC)}.\hskip 1em plus
  0.5em minus 0.4em\relax IEEE, 2017, pp. 460--461.

\bibitem{meng2019gastric}
M.~Meng and M.~Kiani, ``Gastric seed: Toward distributed ultrasonically
  interrogated millimeter-sized implants for large-scale gastric
  electrical-wave recording,'' \emph{IEEE Transactions on Circuits and Systems
  II: Express Briefs}, vol.~66, no.~5, pp. 783--787, 2019.

\bibitem{weber2018miniaturized}
M.~J. Weber, Y.~Yoshihara, A.~Sawaby, J.~Charthad, T.~C. Chang, and
  A.~Arbabian, ``A miniaturized single-transducer implantable pressure sensor
  with time-multiplexed ultrasonic data and power links,'' \emph{IEEE Journal
  of Solid-State Circuits}, vol.~53, no.~4, pp. 1089--1101, 2018.

\bibitem{lee2005logic}
D.-H. LEE, ``Logic design of orthogonal variable spreading factor code
  generator,'' \emph{Journal of Circuits, Systems, and Computers}, vol.~14,
  no.~03, pp. 507--513, 2005.

\bibitem{rowe1964amplitude}
H.~E. Rowe, ``Amplitude modulation with a noise carrier,'' \emph{Proceedings of
  the IEEE}, vol.~52, no.~4, pp. 389--395, 1964.

\bibitem{krimholtz1970new}
R.~Krimholtz, D.~A. Leedom, and G.~L. Matthaei, ``New equivalent circuits for
  elementary piezoelectric transducers,'' \emph{Electronics Letters}, vol.~6,
  no.~13, pp. 398--399, 1970.

\bibitem{redwood1961transient}
M.~Redwood, ``Transient performance of a piezoelectric transducer,'' \emph{The
  journal of the acoustical society of America}, vol.~33, no.~4, pp. 527--536,
  1961.

\bibitem{kino1987acoustic}
G.~S. Kino, \emph{Acoustic waves: devices, imaging and analog signal
  processing}, 1987, no. 43 KIN.

\bibitem{johnson2017implantable}
B.~C. Johnson, S.~Gambini, I.~Izyumin, A.~Moin, A.~Zhou, G.~Alexandrov, S.~R.
  Santacruz, J.~M. Rabaey, J.~M. Carmena, and R.~Muller, ``An implantable
  700$\mu$w 64-channel neuromodulation ic for simultaneous recording and
  stimulation with rapid artifact recovery,'' in \emph{2017 Symposium on VLSI
  Circuits}.\hskip 1em plus 0.5em minus 0.4em\relax IEEE, 2017, pp. C48--C49.

\bibitem{murmann2012thermal}
B.~Murmann, ``Thermal noise in track-and-hold circuits: Analysis and simulation
  techniques,'' \emph{IEEE solid-state circuits magazine}, vol.~4, no.~2, pp.
  46--54, 2012.

\bibitem{bazes1991two}
M.~Bazes, ``Two novel fully complementary self-biased cmos differential
  amplifiers,'' \emph{IEEE Journal of Solid-State Circuits}, vol.~26, no.~2,
  pp. 165--168, 1991.

\bibitem{loeb1977parylene}
G.~E. Loeb, M.~Bak, M.~Salcman, and E.~Schmidt, ``Parylene as a chronically
  stable, reproducible microelectrode insulator,'' \emph{IEEE Transactions on
  Biomedical Engineering}, no.~2, pp. 121--128, 1977.

\end{thebibliography}
\end{document}